\documentstyle[12pt]{article}

\newcommand{\resection}[1]{\setcounter{equation}{0}\section{#1}}
\newcommand{\appsection}[1]{\setcounter{section}{1}
\setcounter{equation}{0}
                  \section*{Appendix: #1}}
\newcommand{\acknowledgements}{\section*{Acknowlegements}}
\renewcommand{\theequation}{\thesection.\arabic{equation}}
\newcommand{\EQ}{\begin{equation}}
\newcommand{\EN}{\end{equation}}
\newcommand{\EQN}{\[}
\newcommand{\ENN}{\]}
\newcommand{\EQA}{\begin{eqnarray}}
\newcommand{\ENA}{\end{eqnarray}}
\newcommand{\spz}{\hspace{0.7cm}}
\newcommand{\hs}{\hspace{0.1cm}}
\newcommand{\deriv}[1]{\frac{\partial\hs}{\partial #1}}
\newcommand{\vtp}{\vartheta(p)}
\newcommand{\vtq}{\vartheta(q)}

\newcommand{\s}{\sigma}
\newcommand{\sm}{\sigma^{(m)}}
\newcommand{\sn}{\sigma^{(n)}}
\newcommand{\snu}{\sigma^{(n_1)}}
\newcommand{\snd}{\sigma^{(n_2)}}
\newcommand{\eps}{\varepsilon}
\newcommand{\adet}{{\rm det}}
\newcommand{\expaq}{\exp\left[\alpha Q(m) \right]}
\newcommand{\exppp}{\exp\left[imp+4it\cos p\right]}
\newcommand{\exppq}{\exp\left[imq+4it\cos q\right]}
\newcommand{\expmp}{\exp\left[-imp-4it\cos p\right]}

\newcommand{\qc}{{\hat q}}
\newcommand{\ic}{{\hat I}}
\newcommand{\vc}{{\hat V}}
\newcommand{\vcs}{{\hat v}}
\newcommand{\rcs}{{\hat r}}
\newcommand{\mc}{{\hat M}}
\newcommand{\gc}{{G}}
\newcommand{\ec}{{E}}
\newcommand{\dc}{{D}}
\newcommand{\rc}{{\hat R}}

\newcommand{\la}{\langle}
\newcommand{\ra}{\rangle}
\newcommand{\psin}{\Psi_{_N}}
\newcommand{\intk}{\int_{-k_F}^{k_F}}
\newcommand{\intp}{\int_{-\pi}^{\pi}}
\newcommand{\goto}{\rightarrow}

\begin{document}
\oddsidemargin 5mm
\setcounter{page}{0}
\renewcommand{\thefootnote}{\fnsymbol{footnote}}
\newpage
\setcounter{page}{0}
\begin{titlepage}
\begin{flushright}
{\small E}N{\large S}{\Large L}{\large A}P{\small P}-L-381/92
\end{flushright}
\begin{flushright}
ITP-SB-92-21
\end{flushright}
\vspace{0.5cm}
\begin{center}
{\large {\bf Determinant Representation for the Time Dependent Correlation
Functions in the XX0 Heisenberg Chain}} \\
\vspace{1cm}
{\bf F. Colomo$^{1}$,
A.G. Izergin$^{2}$
\footnote{Permanent Address: {\em Sankt-Petersburg Branch (POMI) of V.A.
Steklov Mathematical Institute of Russian Academy of Sciences, Fontanka
27, 191011 St.-Petersburg, Russia.}},
V.E. Korepin$^{3}$,
V. Tognetti$^{4}$ \\
\vspace{0.8cm}
$^1${\em I.N.F.N., Sezione di Firenze,
Largo E. Fermi 2, 50125 Firenze, Italy}\\
$^2${\em Forum: Project on Condensed Matter Theory of INFM, Firenze-Pisa,
Italy}\\
and\\
{\em Laboratoire de Physique Th\'eorique ENSLAPP, Ecole Normale
Superieure de Lyon, France}\\
$^3${\em  Institute for Theoretical Physics, State University of
New York at Stony Brook, NY-11794-3840, USA}\\
$^4${\em Dipartimento di Fisica, Universit\`a di Firenze, Largo E. Fermi
2, 50125 Firenze, Italy}}
\end{center}
\vspace{6mm}
\begin{abstract}
Time dependent correlation functions in the Heisenberg XX0 chain in
the external transverse magnetic field are calculated. For a finite chain
normalized mean values of local spin products are represented as
determinants
of $N\times N$ matrices, $N$ being the number of quasiparticles in the
corresponding eigenstate of the Hamiltonian. In the thermodynamical limit
(infinitely long chain), correlation functions are expressed in terms of
Fredholm determinants of linear integral operators.
\end{abstract}
\vspace{3mm}
\centerline{April 1992}
\centerline{REVISED VERSION}
\end{titlepage}

\newpage
\renewcommand{\thefootnote}{\arabic{footnote}}
\setcounter{footnote}{0}

\resection{Introduction}

Recently essential developments have been made in the theory of quantum
correlation functions showing that correlators of quantum exactly solvable
models satisfy classical completely integrable differential equations
\cite{twelve}-\cite{novok} (for the case of the nonrelativistic Bose gas,
this program is fulfilled, see, for instance, \cite{book}). In order to
obtain these differential equations, an important preliminary step consists
in representing correlation functions as the determinants of Fredholm
linear
integral operators. For the nonrelativistic Bose gas these representations
were
given both for the time independent case \cite{eight,nine} and for the
time dependent one \cite{ten}.

In this paper determinant representations of this kind are obtained for
the distance, time and temperature dependent two-point correlation
functions
of the XX0 Heisenberg chain, both for the finite lattice and in the
thermodynamical limit. In order to write differential equations and to
calculate their asymptotics, our further plan is to construct and solve a
matrix Riemann problem, similarly to the case of the nonrelativistic Bose
gas \cite{fourteen}, \cite{IIKtwo}-\cite{sixteen} (see also Ref.
\cite{book}).

The XX0 chain is the isotropic case of the XY model \cite{one}, being also
the
free fermions point for the XXZ chain. The Hamiltonian describing the
nearest
neighbour interaction of local spins $\frac{1}{2}$  situated at the sites
of
the one-dimensional lattice in a constant transverse magnetic field $h$ is
\EQ
H(h)=-\sum_{m=1}^M\left[\s_x^{(m)}\s_x^{(m+1)}+\s_y^{(m)}\s_y^{(m+1)}+h\sm_z
\right]\hs.\label{ham}
\EN
The total number $M$ of sites is supposed to be even and periodical
boundary
conditions, $\s^{(M+1)}_s=\s_s^{(1)}$ ($s=x, y, z$), are imposed. Pauli
matrices are normalized as $\left(\sm_s\right)^2=1$.

The ferromagnetic state $\mid 0\ra\equiv\otimes_{m=1}^{M}\mid\uparrow\ra_m$
(all spins up) is an eigenstate of the Hamiltonian. All the other
$2^M-1$ eigenstates can be constructed by filling this ferromagnetic state
with $N$ quasiparticles ($N=1, 2, ... , M$) possessing quasimomenta $p_a$
($a=1, 2, ... , N$) and energies $\eps(p_a)$,
\EQ
\eps(p)\equiv\eps(p,h)=-4\cos p +2h\hs.\label{energy}
\EN
Due to periodical boundary conditions, one has the following condition
for the permitted values of quasimomenta
\EQ
e^{iMp_a}=(-1)^{N+1}\hs, \hspace{1.5cm} a=1, ... , N\hs.\label{bethe}
\EN
All the  momenta of the quasiparticles in a given eigenstate should be
different, so that, $e.g.$, for $N=M$, one gets in fact only one eigenstate
(which is just the other ferromagnetic state with all spins down,
$\mid 0'\ra=\otimes_{m=1}^M\mid \downarrow\ra_m$).

Due to the similarity transformation,
\EQA
&&H(h)\hs \goto\hs H(-h) = UH(h)U^{-1};\hspace{2.0cm} U=\prod_{m=1}^{M}
\sm_x\hs,\nonumber\\
&&\mid 0 \ra\hs\goto\hs\mid 0'\ra=U\mid 0 \ra\hs,\label{simil1}
\ENA
it is sufficient to consider only nonnegative magnetic fields, $h\geq 0$.
Furthermore the choice of  the minus sign at the r.h.s.
of eq. (\ref{ham}) is just a matter of convenience due to the property
\EQN
H(h)\hs \goto\hs -H(-h) = VH(h)V^{-1};\hspace{2.0cm} V=\prod_{m=1}^{M/2}
\s^{(2m)}_z\hs.\label{simil2}
\ENN

The model in the thermodynamic limit ($M$ $\goto$ $\infty$, $h$ fixed) is
the most interesting. For $h\geq h_c\equiv 2$, the ground state of the
Hamiltonian is just the ferromagnetic state $\mid 0 \rangle$.
For magnetic field smaller than the critical value, $0\leq h< h_c$,
the ground state $\mid\Omega\ra$ is obtained by filling the ferromagnetic
state with quasiparticles
possessing all the allowed values of momenta inside the Fermi zone,
$-k_F\leq p_a\leq k_F$, where the Fermi
momentum $k_F$ is defined by the requirement $\eps(k_F)=0$:
\EQ
k_F=\arccos\left(\frac{h}{2}\right)\hspace{1.5cm}0\leq
h<h_c\hs.\label{fermi}
\EN
In the thermodynamical limit ($M\goto \infty$) the number $N_0$ of
quasiparticles in the ground state is going to infinity, $N_0\goto \infty$,
``density'' $D\equiv N_0/M$ remaining fixed.
At zero magnetic field $k_F=\frac{\pi}{2}$, and there are $\frac{M}{2}$
quasiparticles in the ground state, magnetization being equal to zero.

At non zero temperatures $T>0$, the distribution of quasiparticles in the
momentum space is $\vtp/2\pi$ where $\vtp$ is the Fermi weight
\EQ
\vtp\equiv\vartheta(p,h,T)=\frac{1}{1+\exp\left[\frac{\eps(p)}{T}\right]}\hs
{}.
\label{weight}
\EN
Temperature and time dependent correlators of local  spins $\sm_s(t)$,
\EQA
\sm_s(t)&\equiv&e^{iHt}\sm_s e^{-iHt}\hs,\nonumber\\
\sm_s&\equiv&\sm_s(0)\hs,\hspace{1.5cm} s=x, y, z, \label{time}
\ENA
are defined as usual,
\EQA
g_{sr}^{(T)}(m,t)&\equiv&\la\snd_s(t_2)\snu_r(t_1)\ra_{_T}=\nonumber\\
&=&\frac{{\rm Sp}\left\{\exp[-H/T]\snd_s(t_2)
\snu_r(t_1) \right\}}{{\rm Sp}\left\{\exp[-H/T]
\right\}}\hs.\label{fintemp}
\ENA
Due to translation invariance, the correlators depend only on differences,
\EQ
m=n_2-n_1,\hspace{2cm} t=t_2-t_1\hs.\label{relative}
\EN
At zero temperature, only the ground state $\mid\Omega\ra$
contributes to the traces in (\ref{fintemp}):
\EQ
g^{(0)}_{sr}(m,t)\equiv\frac{\la\Omega\mid\snd_s(t_2)\snu_r(t_1)\mid\Omega
\ra}{\la\Omega\mid\Omega\ra}\hs.\label{zerotemp}
\EN

In Ref. \cite{one} the equal-time correlators ($t=0$) of XY model (of which
the XX0 model is a particular case) were calculated at zero magnetic field
($h=0$). The simple answer for the correlator $g_{zz}^{(T)}(m,0)$ of the
third
spin components was given; for the XX0 chain it reduces in essential to the
square modulus of the Fourier transform  of the Fermi weight. This  result
was generalized to the case of the time-dependent correlator with nonzero
transverse magnetic field \cite{two}. Properties of correlator
$g_{zz}^{(T)}$
were considered in much detail \cite{one}-\cite{four}. Real systems for
experimental comparisons were found \cite{five}.

On the contrary, correlators of the other local spin components are much
more complicated. In Ref. \cite{one} these correlators (for the XY model
at $t=0$, $h=0$) were represented as the determinants of $m\times m$
matrices
($m$ is the distance between correlating spins). This representation was
investigated in detail in \cite{three} (see also \cite{six}). In Ref.
\cite{seven} the structure of the time-dependent correlators was
investigated
on the basis of an extension of the thermodynamic Wick theorem. In Ref.
\cite{sevenprime}, a representation (different from the one obtained
below in this particular case) of the autocorrelator ($m=0$, $t\not=0$)
in the transverse Ising chain in critical magnetic field (closely related
to
correlators in the XX0 model at $h=0$) were given as the Fredholm
determinant
of a linear integral operator. This representation was used in \cite{novok}
to produce differential equations for the autocorrelator.

In the present paper the time dependent correlators (see (\ref{fintemp}),
(\ref{relative}) for notations; $\sm_{\pm}$ $\equiv$
$\frac{1}{2}[\sm_x\pm i\sm_y]$)
\EQA
g_+^{(T)}(m,t)&\equiv&\la\snd_+(t_2)\snu_-(t_1)\ra_{_T}\hs,\label{corrp}\\
g_-^{(T)}(m,t)&\equiv&\la\snd_-(t_2)\snu_+(t_1)\ra_{_T}\hs,\label{corrm}
\ENA
for the XX0 model in a transverse magnetic field are given (in the
thermodynamical limit) as Fredholm determinants of linear integral
operators.
These representations, different from those of paper \cite{one}, are
similar to the representations of two-point correlators previously
obtained for  the
one-dimensional impenetrable Bose gas (see \cite{eight,nine} in the
equal-time
case and \cite{ten} for time dependent correlators). In Ref.
\cite{itoyama},
the Fredholm determinant representation for the time-independent generating
functional of
currents in the sine-Gordon model at the free-fermion point was obtained.
Representations of this kind proved to be extremely useful in obtaining
the integrable differential equations for correlation functions
in the case of the impenetrable Bose gas (the V Painlev\'e transcendent
in the equal time zero temperature case \cite{twelve} and integrable
partial
differential equations for time and temperature dependent correlation
functions
\cite{thirteen,fourteen}). In turn, this fact allowed to construct exact
asymptotics for the correlators \cite{twelve,fifteen,sixteen}.
Corresponding
results are expected to be obtained also for the XX0 chain.

The further contents of this paper is as follows.
In Section 2 the detailed description of the XX0 model on the finite
lattice
is given. In Section 3 the mean value of the generating functional
of the equal-time third spin components correlators with respect to any
eigenfunction of the Hamiltonian is calculated. It is represented as the
determinant of a $N\times N$ matrix, $N$ being the number of quasiparticles
in the corresponding eigenstate. The mean value of the time-dependent
product
of two third spin components on a finite lattice is calculated in Section
4.
In Section 5, form factors of operators $\sm_{\pm}(t)$ ($i.e.$ their matrix
elements between eigenstates of the Hamiltonian) are represented as
determinants
of $N\times N$ matrices. The representation of the normalized mean values
of products $\snd_+(t_2)\snu_-(t_1)$, $\snd_-(t_2)\snu_+(t_1)$, with
respect to any eigenstate of the Hamiltonian containing $N$ quasiparticles,
as the determinants of $N\times N$ matrices (for finite $M$) are given in
Section 6. In Section 7, these representations are proved. The answers for
the correlators in the thermodynamical limit ($M\goto\infty$,  $h$ fixed)
are
given in Section 8 (for zero temperature) and in Section 9 (for non zero
temperature). Some details in performing the thermodynamical limit are
considered in the Appendix.

\resection{The Model on the Finite Lattice}
The Hamiltonian describing the XX0 model in transverse magnetic field
was given in (\ref{ham}):
\EQ
H(h)=H_0-2hS_z\hs.\label{ham2}
\EN
Here $H_0$ describes the nearest neighbour interaction of spins
$\frac{1}{2}$
situated at the sites of the lattice,
\EQ
H_0=-\sum_{m=1}^{M}[\s_x^{(m)}\s_x^{(m+1)}+\s_y^{(m)}\s_y^{(m+1)}]
\hs,\label{hamzero}
\EN
and $S_z$ is the third spin component of the total spin,
\EQ
S_z=\frac{1}{2}\sum_{m=1}^{M} \sm_z\hs.\label{totspin}
\EN
The space ${\cal H}$ where these operators act is a tensor product of
local spaces, ${\cal H}=\otimes_{m=1}^{M}\hs{\cal H}_{(m)}$,
${\cal H}_{(m)}$ being the linear space ${\bf C}^2$ corresponding to the
$m^{th}$ local spin spanned by the basis vectors
$\mid \uparrow\rangle_m$ (spin up), $\mid \downarrow\rangle_m$ (spin down)
\EQA
\sm_z\mid\uparrow\rangle_m=\mid\uparrow\rangle_m\hs; \hspace{1cm} & &
\sm_z\mid\downarrow\rangle_m=\mid\downarrow\rangle_m\hs; \nonumber\\
_m\langle\uparrow\mid\uparrow\rangle_m=\hs _m\langle\downarrow\mid
\downarrow\rangle_m=1\hs;\hspace{7mm} & & _m\langle\uparrow\mid\downarrow
\rangle_m=0\hs. \label{space}
\ENA
Local spin operators $\sm_p$ ($p=x,y,z$) with commutation relations
\EQ
\left[\sm_p,
\s^{(n)}_q\right]=2i\delta_{mn}\epsilon_{pqr}\sm_r\label{pauli}
\EN
are Pauli matrices acting non trivially in ${\cal H}_{(m)}$.
The complete set of mutual eigenstates of $H(h)$ ($H_0$)
and $S_z$ is obtained
by applying lowering operators $\sm_-$ (as usual, $\sm_{\pm}=
\frac{1}{2}\left[\sm_x\pm i\sm_y\right]$) to the ferromagnetic state
$\mid 0 \rangle$ (all spins up):
\EQA
&&\mid 0 \rangle=\otimes_{m=1}^{M}\mid\uparrow\rangle_m,\hs;\nonumber\\
&&S_{z}\mid 0 \rangle=\frac{M}{2} \mid 0 \rangle\hs;\hspace{1cm}
H\mid 0 \rangle=-Mh\mid 0\ra\hs.
\ENA
The explicit form for the eigenfunctions of Hamiltonian (\ref{ham2}) is
well
known, being just the simplest case of eigenfunctions  of the XXZ model
\cite{eleven} with the two-particle scattering phases equal to zero.
For the XX0 model they have the following form:
\EQA
\mid\Psi_N(\{p\})\rangle&\equiv&\mid\Psi_N(p_1, ...,p_N)\rangle=
\label{wave}\\
&=&\frac{1}{\sqrt{N!}}\sum_{m_1,...,m_N}^{M}\chi_{_N}(\{m\}\mid\{p\})\hs
\s^{(m_1)}_- ... \s^{(m_N)}_-\mid 0 \rangle\hs,\nonumber
\ENA
with wave function $\chi_{_N}$ given as
\EQA
\chi_{_N}(\{m\}\mid\{p\})&\equiv&\chi_{_N}(m_1,...,m_N\mid p_1,...,p_N)=
\nonumber\\
&=&\frac{1}{\sqrt{N!}}\left[\prod_{1\leq a<b\leq N}\epsilon(m_b-m_a)
\right]
\cdot\sum_{Q} (-1)^{[Q]}\exp\left[ i\sum_{a=1}^{N}m_a p_{Q_a}\right]\hs,
\label{chi}
\ENA
where $\epsilon(m)$ is the sign-function, defined as
\EQ
\epsilon(m)=\left\{\begin{array}{ll} 1,&\spz m>0\hs;\\
-1,&\spz m<0\hs;\\ 0,&\spz m=0\hs.\end{array}\right.\label{epsilon}
\EN
The sum in (\ref{chi}) is taken over all the permutations of $N$ numbers,
$Q:\hs (1, 2, ... , N)\hs\rightarrow\hs({Q_1}, Q_2, ... , {Q_N})$;
$[Q]$ denotes the parity of the permutation.

Due to the periodical boundary conditions, quasimomenta $p_a$ ($-\pi< p_a
\leq\pi$) satisfy equations
\EQ
\exp\left[i p_a M\right]=(-1)^{N+1}\hs, \spz a=1, ... ,
N\hs,\label{betheone}
\EN
$i.e.$, the permitted values for the momenta are
\EQA
&&\spz p_a=\frac{2\pi}{M}n_a\hs,\nonumber\\
&&\begin{array}{ll}
n_a=-\frac{M}{2}+j\hs,& \spz  j=1,2,...,M\spz {\rm for}\hs N \hs {\rm odd,}
\\
n_a=-\frac{M+1}{2}+j\hs,& \spz  j=1,2,...,M\spz {\rm for}\hs N \hs
{\rm even.}
\end{array}\label{allowed}
\ENA

The sum over the permutations in (\ref{chi}) is just the Slater
determinant.
Hence wave function $\chi_{_N}$ is symmetric in ``coordinates'' $m_a$ and
antisymmetric in quasimomenta $p_a$, being
equal to zero if some of the coordinates or momenta coincide. In
particular, all the momenta $p_a$ ($a=1,...,N$) of the quasiparticles in
the
eigenstate $\mid\Psi_{_N}(\{p\})\rangle$ should be different, otherwise the
wave function is identically equal to zero (``Pauli principle''). $E.g.$,
for $N=M$ only one eigenstate
$\mid\Psi_M\rangle$ $\equiv$ $\mid 0'\rangle$ $=$ $\otimes_{m=1}^{M}
\mid\downarrow\rangle_m$ does exist, which is just the other ferromagnetic
state with all spins down.

Eigenvalues of operators $H(h)$ and $S_z$ for eigenstate
$\mid\Psi_{_N}(\{p\})\rangle$ are
\EQA
H\mid\Psi_N(\{p\})\rangle &=& \left(\sum_{a=1}^N\varepsilon(p_a) \right)
\mid\Psi_N(\{p\})\rangle\hs, \nonumber\\
S_z\mid\Psi_N(\{p\})\rangle&=&\left(\frac{M}{2}-N\right)
\mid\Psi_N(\{p\})\rangle\hs,
\ENA
with the one particle energy (``dispersion law'') given as
\EQ
\varepsilon(p)=-4\cos p + 2h\hs.
\EN

Eigenvectors (\ref{wave}) with different numbers of quasiparticles
are orthogonal,
\EQN
\langle\Psi_{N_1}\mid\Psi_{N_2}\rangle=0\hs, \spz N_1\not=N_2\hs,
\ENN
as well as eigenvectors with the same number of particles, but
different sets of momenta:
\EQN
\langle\Psi_{N}(\{p\})\mid\Psi_{N}(\{ p\})\rangle=0\hs,
\spz\{p\}\not=\{p'\}
\hs;
\ENN
(more exactly, this is valid if the set $\{p'\}$ cannot be obtained from
the set $\{p\}$ by means of permutations of quasimomenta $p_a$).
The normalization is given as
\EQA
\langle\Psi_{N}(\{p\})&\mid&\Psi_{N}(\{ p'\})\rangle=\nonumber\\
&=&\sum_{m_1,...,m_N=1}^{M}\chi^{\ast}_{_N}(m_1,...,m_N\mid p_1,...,p_N)
\hs\chi_{_N}(m_1,...,m_N\mid p_1,...,p_N)=\nonumber\\
&=&M^N\hs.\label{norm}
\ENA

Let us discuss the dependence on the external magnetic
field $h$.  We consider the model only
in the case $h\geq 0$ (which is sufficient, due to property
(\ref{simil1})).
For strong magnetic fields, $h\geq h_c\equiv 2$, the ground state of the
Hamiltonian is just the ferromagnetic state $\mid 0 \rangle$, with
normalized mean value $\langle\sm_z\rangle$ (magnetization) equal to
one. For magnetic field smaller
than the critical value, $0\leq h< h_c$, the ferromagnetic state is not
the ground state. The ground state $\mid \Omega\ra$ in this case is
constructed
by filling the ferromagnetic state with quasiparticles occupying all the
permitted vacancies, see (\ref{allowed}), in the Fermi zone,
$-k_F\leq p_a\leq k_F$, where $k_F$ $=$ $\arccos(h/2)$ is the Fermi
momentum (\ref{fermi}).

Magnetization $\s_{_N}$ in the arbitrary
eigenstate $\mid\Psi_{_N}(\{p\})\rangle$ is easily computed to be
\EQ
\s_{_N}\equiv\langle\sm_z\rangle_{_N}=\frac{2}{M}\la S_z\ra_{_N}=
\frac{\langle\Psi_N(\{p\})\mid
\sm_z\mid\Psi_N(\{p\})\rangle}{\langle\Psi_N(\{p\})\mid
\Psi_N(\{p\})\rangle}=1-\frac{2N}{M}\hs.\label{magnet}
\EN

Let us conclude this Section by discussing the correspondence
between the XX0 model and free fermions on a one-dimensional lattice
\cite{one}. Introducing operator ${\hat q}_n$ of the number
of quasiparticles at the $n^{th}$ site of the lattice,
\EQ
{\hat q}_n=\frac{1}{2}\left(1-\s_z^{(n)}\right)=\s^{(n)}_- \s^{(n)}_+\hs,
\spz\exp\left[2\pi i {\hat q}_n \right]=1\hs,\label{defqsmall}
\EN
and the operator $Q(m)$ of the number of quasiparticles in
the first $m$ sites,
\EQ
Q(m)\equiv\sum_{n=1}^{m}{\hat q}_n\hs;\spz\exp\left[2\pi i Q(m)\right]=1
\hs,\label{Q}
\EN
one constructs fermionic fields $\psi(m)$, $\psi^{\dagger}(m)$ through
the Jordan-Wigner transformation \cite{jordan}:
\EQA
\psi(m)&=&\exp\left[i\pi Q(m)\right]\hs\sm_+\hs,\nonumber\\
\psi^{\dagger}(m)&=&\sm_-\hs \exp\left[i\pi Q(m)\right]\hs,\label{fermion}
\ENA
with the following anticommutation relations:
\EQA
\left\{\psi(m),\psi^{\dagger}(n) \right\}&\equiv&\psi (m)\psi^{\dagger}(n)
+\psi^{\dagger}(n)\psi(m)\hs=\hs\delta_{mn}\hs,\nonumber\\
\left\{\psi(m),\psi(n)
\right\}&=&\left\{\psi^{\dagger}(m),\psi^{\dagger}(n)
\right\}\hs=\hs 0\hs. \nonumber
\ENA
Operators ${\hat q}_n$, $Q(m)$ and $\sm_{\pm}$ may be rewritten in terms
of the fermionic fields as
\EQ
{\hat q}_n=\psi^{\dagger}(n)\psi(n)\hs;\spz
Q(m)=\sum_{n=1}^{m}\psi^{\dagger}
(n)\psi(n)\hs,
\EN
and
\EQA
\sm_+&=&\exp\left[i\pi Q(m)\right]\hs\psi(m)\hs,\nonumber\\
\sm_-&=&\psi^{\dagger}(m)\hs\exp\left[i\pi Q(m)\right]\hs.\label{tailed}
\ENA
It is therefore possible to express the Hamiltonian $H_0$ (\ref{hamzero})
and the total spin $S_z$ (\ref{totspin}) in terms of the fermionic fields:
\EQA
H_0&=&-2\sum_{m=1}^{M}\left[\psi^{\dagger}(m+1)\psi(m)+\psi^{\dagger}(m)
\psi(m+1)\right]\hs,\nonumber\\
S_z&=&\frac{M}{2}-\sum_{m=1}^{M}\psi^{\dagger}(m)\psi(m)\hs,
\ENA
which indeed  describe free fermions on the lattice \cite{one}.
Hence the problem of computing correlation functions of local operators,
(as $e.g.$ $\langle\s^{(n_2)}_+(t_2)\s^{(n_1)}_-(t_1)\rangle$) in the XX0
model is equivalent to calculating correlators of ``tailed''
(``disordered'')
operators (\ref{tailed}) for free fermions.

 Our calculations below are done directly in the frame
of the XX0 model itself.

\resection{Normalized Mean Value of Operator $\exp\left[\alpha Q(m)
\right]$
on the Finite Lattice}

Let us consider the normalized mean value
\vspace{2mm}
\EQ
\langle\exp\left[\alpha Q(m) \right]\rangle_{_N}\equiv
\frac{\langle\Psi_N(\{p\})
\mid\exp\left[\alpha Q(m) \right]\mid\Psi_N(\{p\})\rangle}{\langle
\Psi_N(\{p\})\mid \Psi_N(\{p\})\rangle}\hs.\label{mean1}
\EN
\vspace{2mm}\noindent
in the XX0 model on the finite periodical lattice with $M$ sites.
Here $Q(m)$ (\ref{Q}) is the operator of the number of quasiparticles
in the first $m$ sites of the lattice; $\alpha$
is a complex parameter. The mean value (\ref{mean1}) is taken with respect
to some eigenstate $\mid\Psi_N(\{p\})\rangle$ (\ref{wave}) of Hamiltonian
(\ref{ham2}). As explained in the end of this Section, the quantity
$\exp[\alpha Q(m)]$ defined in (\ref{mean1}) generates equal-time mean
values of products of operators $\sm_z$; so it is called
``generating functional'' for these mean values.
The value of generating functional (\ref{mean1}) itself  at
$\alpha=-\infty$ has a clear physical meaning giving the probability that
there are no quasiparticles  in the first $m$ sites of the lattice in
the eigenstate $\mid\Psi_{_N}(\{p\})\ra$.

The generating functional can be represented in the following explicit
form:
\EQ
\langle\exp\left[ \alpha Q(m)\right]\rangle_{_N}=\adet_{_N} {\cal M}(m)
\hs.\label{gener}
\EN
Here the r.h.s. is the determinant of the $N\times N$
matrix ${\cal M}$ $\equiv$ ${\cal M}(m)$ with matrix elements given by
\EQ
\left({\cal M}\right)_{ab}=\delta_{ab}\left(1+\frac{e^{\alpha}-1}{M}
m\right)
+(1-\delta_{ab})\frac{e^{\alpha}-1}{M}\cdot\frac{
\sin\frac{m}{2}(p_a-p_b)}{\sin\frac{1}{2}(p_a-p_b)}\hs,\spz a,b=1,2,...,N
\hs,\label{tretre}
\EN
where $\{p\}=\{p_1,...,p_N\}$, $p_a=\frac{2\pi}{M}n_a$ is the set of
quasimomenta (\ref{allowed}) defining eigenstate $\mid\Psi(\{p\})\rangle$.

Let us explain how this expression for the generating functional is
obtained. First, using commutation relation
\EQA
& &\expaq\hs\s^{(n)}_-=\varphi(n,m)\s^{(n)}_-\hs\expaq\hs,\nonumber\\
& &\varphi(n,m)=\left\{\begin{array}{ll}
1, & \spz n>m, \\ e^{\alpha}, & \spz n\leq m,\end{array}\right.
\ENA
and also equations (\ref{wave}), (\ref{space}), as well as commutation
relations (\ref{pauli}) between the Pauli matrices, one gets
\EQA
\langle\expaq\rangle_{_N}&=&\frac{1}{M^N} \sum_{m_1,...,m_N}^{M}
\chi^{\ast}(m_1,...,m_N\mid
p_1,...,p_N)\hs\chi(m_1,...,m_N\mid p_1,...,p_N)\cdot\nonumber\\
& &\hspace{2cm}\cdot\varphi(m_1,m)\hs\varphi(m_2,m)
\hs...\hs\varphi(m_N,m)\hs.
\ENA

Now one uses equation (\ref{chi}) for $\chi_{_N}$, and performs explicitly
the summations over $m_1,...,m_N$ by means of formula
\EQN
\sum_{n=1}^Me^{i n (p_a-p_b)}\varphi(n,m)=\left(M+(e^{\alpha}-1)m\right)
\delta_{ab}+(1-\delta_{ab})(e^{\alpha}-1)\frac{e^{im(p_a-p_b)}-1}{1
-e^{-i(p_a-p_b)}}
\ENN
with the result
\EQ
\langle\expaq\rangle_{_N}=\frac{1}{N!}\sum_{Q,Q'} (-1)^{[Q]+[Q']}
\prod_{a=1}^{N}\left({\tilde{\cal M}}\right)_{Q_a Q'_a}=
\adet_{_N}{\tilde{\cal M}}\hs,
\EN
where the sum is taken over permutations $Q$, $Q'$ of $N$ numbers. Matrix
elements $({\tilde{\cal M}})_{ab}$ are
\EQA
\left({\tilde{\cal M}}\right)_{ab}&=&\delta_{ab}\left(1+
\frac{e^{\alpha}-1}{M}m \right)+\nonumber\\
&&\spz+(1-\delta_{ab})\frac{e^{\alpha}-1}{M}
\cdot\frac{\exp\left[i m (p_a-p_b)\right]-1}{1-\exp\left[-i
(p_a-p_b)\right]}
\hs,\nonumber\\
&&\hspace{2cm} a,b=1, ... , N.\nonumber
\ENA
Now it is easy to see that the determinants of matrices ${\tilde{\cal M}}$
and ${\cal M}$ (\ref{tretre}) are equal, the matrices being similar.
So one comes to representation (\ref{gener})
for the generating functional.

Normalized mean values for operators $Q(m)$ and ${\hat q}_m$ can be derived
from the generating functional (\ref{mean1}) as follows:
\EQA
\langle Q(m)\rangle_{_N}&=&\left.\deriv{\alpha}\langle
\expaq\rangle_{_N}\right\vert_{\alpha=0}=\frac{N m}{M}\hs;\nonumber\\
\langle {\hat q}_m\rangle_{_N}&=&{\cal D}_1\langle Q(m)\rangle_{_N}=
\frac{N}{M}\hs,
\ENA
where ${\cal D}_1$ is the ``first lattice derivative'' acting on functions
$f(m)$ as $({\cal D}_1 f)(m)\equiv f(m)-f(m-1)$. Relation (\ref{magnet})
is obviously reproduced for magnetization:
\EQ
\s_{_N}\equiv\langle\sm_z\rangle_{_N}=1-2\langle {\hat q}_m\rangle_{_N}=
1-\frac{2N}{M}\hs.
\EN

For the mean value of operator $Q^2(m)$ we readily get
\EQA
\langle Q^2(m)\rangle_{_N}&=&\frac{\partial^2\hs}{\partial\alpha^2}\left.
\langle\expaq\rangle_{_N}\right\vert_{\alpha=0}\hs= \nonumber\\
&=&\frac{N(N-1)}{M^2} m^2+\frac{N m}{M}-\frac{1}{M^2}
\sum_{\begin{array}{c}\\ [-12mm] _{a,b=1}\\ [-5mm]
_{a\not=b}\end{array}}^N
\frac{\sin^2\frac{m}{2}(p_a-p_b)}{\sin^2
\frac{1}{2}(p_a-p_b)}\hs.\label{qqq}
\ENA

Using translation invariance one expresses normalized mean values
$\langle \qc_{m+1} \qc_1 \rangle_{_N}$ = $\langle\qc_{n_2}\qc_{n_1}
\rangle_{_N}$ ($m\equiv n_2-n_1$) as
\EQA
\langle\qc_{n_2} \qc_{n_1}\rangle_{_N}&=& \frac{1}{2}{\cal D}_2\langle
Q^2(m) \rangle_{_N}\hs,\spz m\geq 2\hs,\nonumber\\
\langle\qc_{n+1} \qc_n\rangle_{_N}&=&\frac{1}{2} \langle
Q^2(m=2)\rangle_{_N}
-\langle\qc_n\rangle_{_N}\hs, \nonumber\\
\langle\qc^2_{n}\rangle_{_N}&=& \langle \qc_n\rangle_{_N}\hs,\nonumber
\ENA
where ${\cal D}_2$ is the second derivative on
the lattice acting on $f(m)$ as $({\cal D}_2 f)(m)$ = $f(m+1)+f(m-1)-2
f(m)$.
Using relation (\ref{defqsmall}), ${\hat q}_m=\frac{1}{2}(1-\sm_z)$ , one
obtains for the normalized mean value of the third spin components from
(\ref{qqq}):
 \EQ
\langle\s_z^{(n_2)}\s_z^{(n_1)}\rangle_{_N}=\s_{_N}^2-\frac{4}{M^2}\left\ver
t
\sum_{a=1}^{N}e^{imp_a}\right\vert^2\hs,\spz m\equiv
n_2-n_1\not= 0\hs. \label{corrzfin}
\EN

\resection{Time-Dependent Normalized Mean Value of Operator
$\s_z$ $\s_z$ on the Finite Lattice}

We consider here the simplest time-dependent two-point normalized mean
value,
\EQ
\langle\s^{(n_2)}_z (t_2)\s^{(n_1)}_z(t_1)\rangle_{_N}\equiv
\frac{\langle\Psi_N(\{p\})
\mid\s^{(n_2)}_z (t_2)\s^{(n_1)}_z(t_1)\mid\Psi_N(\{p\})\rangle}{\langle
\Psi_N(\{p\})\mid \Psi_N(\{p\})\rangle}.\label{meanz}
\EN
Here
\EQA
\s_z^{(n)}(t)&=&\exp\left[i H t \right]\s^{(n)}_z(0)\exp\left[-i H t
\right]
\hs, \nonumber\\
\s_z^{(n)}(0)&\equiv&\s^{(n)}_z \hs,
\ENA
is the Heisenberg time-dependent operator of the third spin component
at the $n^{th}$ lattice site; $\mid\Psi_{_N}(\{p\})\ra$ is any
eigenfunction
of the XX0 Hamiltonian (with periodical boundary conditions) parametrized
by
quasimomenta $p_a$ ($a=1, ... , N$), see (\ref{wave})-(\ref{epsilon}).
The result of calculating mean value (\ref{meanz}) is
\EQA
\langle\s^{(n_2)}_z (t_2)\s^{(n_1)}_z(t_1)\rangle_{_N}&=&\s_{_N}^2-
\frac{4}{M^2}\left\vert\sum_{a=1}^N\exp\left[imp_a-it\eps(p_a)
\right] \right\vert^2+ \nonumber\\
&&+\frac{4}{M^2}\left(\sum_{a=1}^{N}\exp\left[-imp_a+it\eps(p_a) \right]
\right)\cdot\left(\sum_{j=1}^{M}\exp\left[imq_j-it\eps(q_j) \right]
\right)\hs,\nonumber\\
&& m\equiv n_2-n_1\hs;\spz t\equiv t_2-t_1\hs.\label{corrzt}
\ENA
Here $\s_{_N}=1-\frac{2N}{M}$  is the magnetization (\ref{magnet}), and
$\eps(p)=-4\cos p +2h$ is the energy (\ref{energy}) of a quasiparticle. The
last sum over $j$ in (\ref{corrzt}) is taken over all the allowed values
\EQ
q_j=\frac{2\pi}{M}\left(-\frac{M}{2}-\frac{1}{4}\left[1+(-1)^N \right]
+j \right)\hs, \spz j=1,...,M\hs,
\EN
of quasimomenta. It should be noted that for $t=0$, $m\not=0$ the sum over
$j$ in (\ref{corrzt}) is equal to zero, so that the equal-time
correlator (\ref{corrzfin}) is reproduced.
As mentioned in Introduction, the $zz$ correlations were already studied in
Ref. \cite{one,two}. Here we present the derivation of eq. (\ref{corrzt})
directly in the frame of the XX0 model.

Let us explain briefly the derivation of formula (\ref{corrzt}).
Inserting in the r.h.s. of eq. (\ref{meanz}) the complete set of
$N$-quasiparticle normalized eigenstates,
\EQ
\frac{\mid\Psi_{_N}(\{q\})\hs\ra\la\hs\Psi_{_N}(\{q\})\mid}{\la
\Psi_{_N}(\{q\})\mid\Psi_{_N}(\{q\})\ra}\hs, \label{complete}
\EN
(since  operators $\s^{(n)}_z$ does not change the number of
quasiparticles,
only states with $N$ quasiparticles do contribute) and taking into account
normalization (\ref{norm}), one gets
\EQA
\langle\s^{(n_2)}_z (t_2)\s^{(n_1)}_z(t_1)\rangle_{_N}&=&
\frac{1}{M^{2N}}\sum_{\{q\}} Z^{\ast}_{_N}(n_2,\{q\},\{p\})\hs
Z_{_N}(n_1,\{q\},\{p\})\cdot\nonumber\\
&&\hspace{1.5cm}\cdot\exp\left\{-it\sum_{a=1}^N\left[\eps(q_a)-\eps(p_a)
\right]\right\}\hs,\label{five}\\
&&\spz t\equiv t_2-t_1.\nonumber
\ENA
Here $Z_{_N}$ is the time-independent form factor of operator
$\sn_z\equiv\sn_z(t=0)$:
\EQ
Z_{_N}(n_1,\{q\},\{p\})\equiv\la\psin(\{q\})\mid\sn_z\mid\psin(\{p\})\ra.
\EN
The sum in (\ref{five}) is taken over all  the
different eigenstates parametrized by different momenta $\{q\}$ $=$
$q_1, ... , q_{_N}$. State $\mid\psin(\{q\})\ra$ is antisymmetric under
permutations of quasimomenta $q_a$'s, and states $\mid\psin(\{q\})\ra$
and $\mid\psin(\{q'\})\ra$ differ at most in the sign if set $\{q'\}$ can
be obtained from set
$\{q\}$ by permutations of momenta $q$. Such $N!$ states are essentially
the
same, and only one of them (anyone, due to the symmetry in $q$'s of
expression
(\ref{complete})) should be inserted as intermediate state.

Using commutation relations (\ref{pauli}) between Pauli matrices, the
symmetry
in arguments $m$ of wave function $\chi_{_N}$ (\ref{chi}), the fact
that $\chi_{_N}=0$ if any two (or
more) $m$'s coincide,  and relation $\sn_z\mid 0\ra$ = $\mid 0 \ra$,
one calculates for the form factor substituting expressions (\ref{wave})
for
the eigenstates:
\EQA
Z_{_N}(n,\{q\},\{p\})&\equiv&\langle\Psi_N(\{q\})\mid\s^{(n)}_z\mid
\Psi_N(\{p\})\rangle=\nonumber\\
&=&\sum_{m_1,...,m_{N}=1}^M\chi^{\ast}_{_N}(m_1,...,m_{N};\{q\})\hs
\chi_{_N}(m_1,...,m_{N};\{p\})\\
&&-2N\sum_{m_1,...,m_{N-1}=1}^M\chi^{\ast}_{_N}(m_1,...,m_{N-1},n;\{q\})
\hs\chi_{_N}(m_1,...,m_{N-1},n;\{p\}).\nonumber
\ENA
Using now the explicit form (\ref{chi}) of the eigenfunctions, one gets
\EQA
Z_{_N}(n,\{q\},\{p\})&=&\frac{1}{N!}\sum_{m_1,...,m_{N}=1}^M\sum_{Q',Q}
(-1)^{[Q]+[Q']}\exp\left[-i\sum_{a=1}^{N} m_a(q_{Q'_a}-p_{Q_a})\right]
-\nonumber\\
&&-\frac{2N}{N!}\sum_{m_1,...,m_{N-1}=1}^M\sum_{Q',Q}
(-1)^{[Q]+[Q']}\exp\left[-in(q_{Q'_N}-p_{Q_N})\right]\cdot\nonumber\\
&&\hspace{2cm}\cdot\exp\left[-i\sum_{a=1}^{N-1} m_a(q_{Q'_a}-p_{Q_a})
\right]\hs.\label{foureight}
\ENA
It is to be mentioned that though, $e.g.$, $\epsilon^2(m_a-m_b)$
= $1-\delta_{m_a,m_b}$ $\not=$ $1$, terms proportional to
$\delta_{m_a,m_b}$
do not contribute, since the sums over permutations in (\ref{foureight})
are equal to zero at $m_a=m_b$, not depending on the value $\epsilon(0)$.
The sums over $m_a$ may be performed explicitly,
\EQ
\sum_{m=1}^M e^{-im(q_a-p_b)}=\left\{\begin{array}{ll}
\hs M,\spz& q_a=p_b,\\ \hs 0,\spz& q_a\not= p_b,\end{array}
\right.
\EN
(periodical boundary conditions (\ref{betheone}), $e^{iMq_a}=e^{iMp_b}=
(-1)^{N+1}$, should be taken into account). It is therefore evident that
the r.h.s. of (\ref{foureight}) does not vanish only  if sets $\{q\}$
and $\{p\}$ differs at most in one momentum. In such cases, we obtain
\EQA
&&Z_{_N}(n;p_1,...,p_{_N};p_1,...,p_{_N})=M^N \left(1-\frac{2N}{M}
 \right)\hs,\label{fourten}\\
&&Z_{_N}(n;p_1,...,p_{_{N-1}},q;p_1,...,p_{_{N-1}},p)=-2M^{N-1}
e^{-in(q-p)}
\hs,\hspace{1.5cm} q\not= p\hs.\label{foureleven}
\ENA
All the non zero form factors, being antisymmetric under permutations  of
$q$'s and permutations of $p$'s, can be easily obtained from
(\ref{fourten}),
(\ref{foureleven}) just by prescribing the corresponding sign.
Turning now to the normalized mean value (\ref{five}) and taking into
account that the product $Z^{\ast}_{_N} Z_{_N}$ is symmetrical under
permutations of $q$'s and of $p$'s, being equal to zero if any two $q$'s
(or any two $p$'s) coincide, one gets
\EQA
&&\la\snd_z(t_2)\snu_z(t_1)\ra_{_N}=\left(1-\frac{2 N}{M}\right)^2+
\label{fourtwelve}\\
&&\hspace{1cm}+\frac{4}{M^2}\sum_{a=1}^N
\sum_{\begin{array}{c}\\ [-12mm] _{j=1}\\ [-5mm]
_{q_j\not=p_1,..., p_N}\end{array}}^M
\exp\left[im(q_j-p_a)-it[\eps(q_j)-\eps(p_a)] \right]\hs.\nonumber
\ENA
Finally, taking into account that
\EQ
\sum_{\begin{array}{c}\\ [-12mm] _{j=1}\\ [-5mm]
_{q_j\not=p_1,..., p_N}\end{array}}^M
f(q_j) =\sum_{j=1}^M f(q_j)-\sum_{a=1}^N f(p_a)\hs,
\EN
one comes immediately to representation (\ref{corrzt}).

\resection{Form factors of Operators $\s_{\pm}$ on the Finite Lattice}

In this Section form factors of operators
$\sm_{\pm}(t)$,
\EQN
\sm_{\pm}(t)=e^{iHt}\hs\sm_{\pm}\hs e^{-iHt}\hs.
\ENN
$i.e.$ their matrix elements between eigenstates of
Hamiltonian (\ref{ham2}) are calculated. This is, in particular, necessary
to calculate the corresponding correlators. Form factors are defined as
\EQA
F_{_N}(m,t,\{q\},\{p\})&\equiv&\langle\Psi_{N+1}(\{q\})\mid\sm_-(t)
\mid\Psi_N(\{p\})\rangle\hs, \label{defform}\\
G_{_N}(m,t,\{p\},\{q\})&\equiv&\langle\Psi_{N}(\{p\})\mid\sm_+(t)
\mid\Psi_{N+1}(\{q\})\rangle\hs.\label{ftwo}
\ENA

The form factors $F_{_N}$, $G_{_N}$ are related by complex conjugation,
\EQ
F^{\ast}_{_N}(m,t,\{q\},\{p\})=G_{_N}(m,t,\{p\},\{q\})\hs,\label{ffour}
\EN
so that in the following only form factor $F_N$ is considered.
In the previous formulae, $\{q\}=q_1,...,q_{_{N+1}}$ (dim$\{q\}=N+1$) and
$\{p\}=p_1,...,p_{_N}$ (dim$\{p\}=N$) are the sets of different
quasimomenta
parametrizing eigenvectors $\mid\Psi_{N+1}\ra$ and $\mid\Psi_{N}\ra$
(if dim$\{q\}\not=$ dim$\{p\}+1$, then the form factors are equal to
zero). It should be emphasized that, due to periodical boundary
conditions (\ref{betheone})
\EQ
e^{iMq_a}=(-1)^N\hs,\hspace{2cm} e^{iMp_b}=(-1)^{N+1}\hs,\label{bethetwo}
\EN
quasimomenta $q_a$ and $p_b$ never coincide.

Form factor $F_{_N}$ (\ref{defform}) can be represented (up to the factor
obvious from translational invariance) as the determinant of a $N\times N$
matrix, namely
\EQA
F_{_N}(m,t,\{q\},\{p\})&=&i^N\exp\left[-im\{\sum_{a=1}^{N+1}q_a-
\sum_{b=1}^{N} p_b\}+it\{\sum_{a=1}^{N+1}\eps(q_a)-\sum_{b=1}^{N}
\eps(p_b)\}\right] \hs\cdot \nonumber\\
&&\spz\cdot{\cal F}_N(\{q \}, \{ p\})\hs, \nonumber\\
{\cal F}_{_N}(\{q\},\{p\})&=&{\cal
F}^{\ast}_{_N}(\{q\},\{p\})\hs,\label{fsix}
\ENA
where function ${\cal F}_N$ does not depend on $m$, $t$, and can be
represented as the determinant
\EQA
{\cal F}_N(\{q\},\{p\})&=&\left.\left(1+\frac{\partial\hs}{\partial z}
\right)\adet_{_N} A(z)\right\vert_{z=0}= \nonumber\\
&=&\adet_{_N} A(z=1)\hs.\label{fseven}
\ENA
Here $A(z)$ is a $N\times N$ matrix with elements depending linearly on the
complex parameter $z$:
\EQA
A(z)&\equiv& A^{(1)}-z A^{(2)}\hs,\nonumber\\
A^{(1)}_{ab}&=& \cot\frac{1}{2}(q_a-p_b)\hs,\label{amat}\\
A^{(2)}_{ab}&=& \cot\frac{1}{2}(q_{_{N+1}}-p_b)\hs,\spz a,b=1,...,N\hs.
\nonumber
\ENA
All the rows of matrix $A^{(2)}$ being the same, the rank of this matrix
is equal to one. Hence, $\adet_{_N} A(z)$ is itself a linear function of
$z$.
For generic linear functions $f(z)$, the following obvious relation holds
\EQ
\left[1+\left.\frac{\partial\hs}{\partial z}\right] f(z)\right\vert_{z=0}=
f(1)\hs,\hspace{2cm} f(z)=a +bz\hs, \label{fnine}
\EN
so that the second equation in (\ref{fseven}) is evident.

Let us explain briefly the derivation of representation (\ref{fseven}),
which is similar to the one of paper \cite{ten} for the nonrelativistic
Bose
gas. Starting from definition (\ref{defform}) and using explicit expression
(\ref{wave}), (\ref{chi}) for the eigenfunctions involved, one gets for the
form factor
\EQA
&&F_{_N}\cdot\exp\left([-it[\sum_{a}^{N+1}\eps(q_a)-\sum_{b=1}^{N}\eps(p_b)]
\right)=\nonumber\\
&&\spz=\sqrt{N+1}\sum_{m_1,...,m_{_N}=1}^{M}\chi^{\ast}_{_{N+1}}
(m_1,...m_{_N},m \mid\{q\})\hs\chi_{_N}(m_1,...,m_N\mid\{p\})=\nonumber\\
&&\spz=\frac{1}{N!}\sum_{m_1,...,m_N=1}^{M}\sum_{Q,P} (-1)^{[Q]+[P]}
\exp\left[-imq_{_{Q_{N+1}}}\right]
\hs\prod_{b=1}^{N}\epsilon(m-m_b)\cdot\nonumber\\
&&\hspace{1.5cm}\cdot\exp\left[-i\sum_{a=1}^N(q_{Q_a}-p_{P_a}) \right]\hs.
\label{intform}
\ENA
Here $Q$: $(1,...,N+1)\goto(Q_1,...,Q_{_{N+1}})$ and $P$: $(1,...,N)\goto
(P_1,...,P_{_N})$ are all the permutations of $N+1$ and $N$ numbers,
respectively. The summations over $m_i$'s are performed explicitly by
means of the formula
\EQ
\sum_{n=1}^M\epsilon(m-n)\hs
e^{-in(q_a-p_b)}=i\cot\frac{1}{2}(q_a-p_b)\cdot
e^{-im(q_a-p_b)}\hs,\label{ften}
\EN
(relation $e^{iM(q_a-p_b)}=-1$, following from periodical boundary
conditions
(\ref{bethetwo}) should be taken into account). Moreover, noticing that the
form factor is
antisymmetric under permutations of quasimomenta $p_b$, one gets
\EQ
F_{_N}=i^N\exp\left[-im[\sum_{a=1}^{N+1}q_a-\sum_{b=1}^N p_b]+
it[\sum_{a=1}^{N+1}\eps(q_a)-\sum_{b=1}^N\eps(p_b)]\right]
\cdot{\cal F}_{_N}(\{q\},\{p\})\hs,
\EN
with
\EQ
{\cal F}_{_N}(\{q\},\{p\})=\sum_{Q} (-1)^{[Q]}\prod_{a=1}^N\cot\frac{1}{2}
(q_{Q_a}-p_a)\hs,\label{fothree}
\EN
(the permutations of $p_a$'s have been summed over, cancelling the factor
$\frac{1}{N!}$ in (\ref{intform})). The sum in the r.h.s. is just the
determinant of the $(N+1)\times(N+1)$ matrix $B$:
\EQA
{\cal F}_N(\{q\},\{p\})&=&\adet_{_{N+1}} B\hs;\\
B_{ab}&=&\cot\frac{1}{2}(q_a-p_b)\hs,\hspace{7mm} a=1,...,N+1, \nonumber\\
B_{a,N+1}&=&1\hs, \hspace{3.4cm} b=1,...,N;\nonumber
\ENA
Subtracting the last row of matrix $B$ from the first N rows (which does
not
change  the value of the determinant) and then expanding the determinant
of the obtained $(N+1)\times (N+1)$ matrix in the elements of the
$(N+1)^{th}$
column, we come exactly to to the determinant of the $N\times N$ matrix
$A(z=1)$ defined in (\ref{amat}):
\EQ
\adet_{_{N+1}} B= \adet_{_N} A(z=1)\hs,
\EN
so that representation (\ref{fseven}) for the form factor is proved.

Let us conclude by mentioning a useful property of representation
(\ref{fseven}): it is possible to introduce a complex parameter into the
r.h.s. of (\ref{fseven}) without changing the result. Let us define
matrix ${\tilde A}(z,c)$ with matrix elements
\EQ
{\tilde A}_{ab}(z,c)\equiv A_{ab}(z)+c(1-z)\hs,\hspace{2cm} c\in{\bf C}\hs.
\EN
It is obvious that $\adet_{_N} {\tilde A}(z,c)$ is a linear function of
$z$,
and moreover, that
\EQ
\adet_{_N} {\tilde A}(z=1,c)=\adet_{_N} A(z=1)\hs,
\EN
so that we can also write for ${\cal F}_N$:
\EQ
{\cal F}_N(\{q\},\{p\})=\left[ 1+\frac{\partial\hs}{\partial z}\right]
\left.\adet_{_N} {\tilde A}(z,c)\right\vert_{z=0}=\adet_{_N}
{\tilde A}(z=1,c)\hs, \label{ftone}
\EN
(the r.h.s does not in fact depend on $c$). The  possibility of introducing
this arbitrary parameter $c$ is related to the possibility of prescribing
any value $\epsilon(0)$ for function $\epsilon(m)$ in (\ref{epsilon}),
(\ref{ften}) without changing the final results of calculations. This is
a consequence of the fact that wave function $\chi$ defined in
(\ref{chi}) is equal to zero if $m_j=m_k$ ($j\not=k$) for any choice of
the value $\epsilon(0)$.

\resection{Time Dependent Normalized Mean Values of Products
of Operators $\s_+$, $\s_-$ on the Finite Lattice}

In this Section the determinant representations for the normalized
mean values of the product of local operators $\sm_+$, $\sn_-$ on the
finite lattice are given. These representations
are proved in the next Section.

Let us consider first the normalized mean value
\vspace{2mm}
\EQ
\la\snd_+(t_2)\snu_-(t_1)\ra_{_N}\equiv\frac{\la\Psi_N(\{p\})\mid
\snd_+(t_2)\snu_-(t_1)\mid\Psi_N(\{p\})\ra}{\la\Psi_N(\{p\})\mid
\Psi_N(\{p\})\ra}\label{sone}
\EN
\vspace{2mm}
with respect to some eigenfunction $\mid\Psi_N(\{p\})\ra$ of the XX0
Hamiltonian
(\ref{ham2}). Due to translational invariance this quantity depends only
on relative distance $m=m_2-m_1$ and time $t=t_2-t_1$. It is also easily
seen that
\EQ
\la\snu_+(t_1)\snd_-(t_2)\ra_{_N}=\la\snd_+(t_2)\snu_-(t_1)\ra_{_N}^{\ast}\h
s,
\EN
so that in the following it is sufficient to consider mean value
(\ref{sone}) in region $m\geq 0$, $-\infty<t<+\infty$.

The following representation is obtained in the next Section
for quantity (\ref{sone}):
\EQA
\la\snd_+(t_2)\snu_-(t_1)\ra_{_N}&=&e^{-2iht}\left[g(m,t)+\deriv{z} \right]
\left.\adet_{_N}\left[S-zR^{(+)} \right]\right\vert_{z=0}=\nonumber\\
&=&e^{-2iht}\left\{\left[g(m,t)-1\right]\adet_{_N} S+\adet_{_N}
\left[S-R^{(+)}\right]\right\}\hs,\label{sthree}
\ENA
Here matrix elements of $N\times N$ matrices $S=S(m,t,\{p\})$
and $R^{(+)}=R^{(+)}(m,t,\{p\})$ are
\EQA
S_{ab}&=&\delta_{ab} \hs d(m,t,p_a)\exp\left[-imp_a-4it\cos p_a\right]+
\nonumber\\
&
&\hs+\left(1-\delta_{ab}\right)\frac{e_+(m,t,p_a)e_-(m,t,p_b)-e_-(m,t,p_a)
e_+(m,t,p_b)}{M\tan\frac{1}{2}(p_a-p_b)}-\nonumber\\
& &\hs-\frac{1}{M}g(m,t)e_-(m,t,p_a)e_-(m,t,p_b)\hs;\label{sfour}\\
R^{(+)}_{ab}&=&\frac{1}{M}e_+(m,t,p_a) e_+(m,t,p_b)\hs.
\ENA
Functions $e_{\pm}$ are defined as
\EQA
e_-(m,t,p_a)&\equiv&\exp\left[-\frac{im}{2}p_a-2it\cos p_a\right]\hs,
\nonumber\\
e_+(m,t,p_a)&\equiv&e_-(m,t,p_a)e(m,t,p_a)\hs,\label{ssix}
\ENA
and functions $g$, $e$, $d$ are given as the sums:
\EQA
g(m,t)&\equiv&\frac{1}{M}\sum_{q}\exp\left[imq+4it\cos q\right]\hs,
\label{sseven}\\
e(m,t,p_a)&\equiv&\frac{1}{M}\sum_{q}\frac{\exp\left[imq+4it\cos q\right]}{
\tan\frac{1}{2}(q-p_a)}\hs,\label{seight}\\
d(m,t,p_a)&\equiv&\frac{1}{M^2}\sum_{q}\frac{\exp\left[imq+4it\cos
q\right]}{
\sin^2\frac{1}{2}(q-p_a)}\hs,\label{snine}
\ENA

Let us explain notations in more detail. Momenta
$p_a$ ($a=1,...,N$) are the momenta of quasiparticles in state
$\mid\Psi_N(\{p\})\ra$ satisfying equations
\EQA
e^{iMp_a}&=&(-1)^{N+1}\hs,\spz -\pi<p_a\leq\pi\hs,\nonumber\\
p_a&=&\left\{\begin{array}{ll}
\frac{2\pi}{M}\left(-\frac{M+1}{2}+n_a \right)\hs,&\spz N \hs{\rm even},
\label{sten}\\
\frac{2\pi}{M}\left(-\frac{M}{2}+n_a \right)\hs,&\spz N \hs{\rm odd},
\end{array} \right.\\
n_a&=&1,...,M\hs,\spz a=1,...,N\hs.\nonumber
\ENA
The sums over $q$'s are taken over all the permitted values of momenta of
quasiparticles in a $(N+1)$-particle state, $i.e$,
\EQ
\sum_{q}f(q)\equiv\sum_{j=1}^{M} f(q_j)\hs,\label{soone}
\EN
where momenta $q_j$ satisfy equations
\EQA
e^{iMq_j}&=&(-1)^{N+2}\hs,\nonumber\\
q_j&=&\left\{\begin{array}{ll}
\frac{2\pi}{M}\left(-\frac{M}{2}+j \right)\hs,&\spz N \hs{\rm even},
\label{sotwo}\\
\frac{2\pi}{M}\left(-\frac{M+1}{2}+j \right)\hs,&\spz N \hs{\rm odd},
\end{array} \right.\\
j&=&1,...,M\hs.\nonumber
\ENA
Due to (\ref{sten}), (\ref{sotwo}), $q_j$ never coincide with any of
$p_a$'s,
so that the sums (\ref{sseven})-(\ref{snine}) are well defined.

The rank of matrix $R^{(+)}$ being  equal to one, the first determinant
in (\ref{sthree}) is a linear function of $z$, and due to property
(\ref{fnine}) the second equality in (\ref{sthree}) is obvious.

Representation (\ref{sthree}) is simplified in the equal-time case
($t=0$), where functions $g$, $e$, $d$ can be computed explicitly:
\EQA
g(m,0)&=&\delta_{m,0}\hs,\hspace{3cm} (m=0,1,...,M-1)\nonumber\\
e(m,0,p_a)&=&i\left(1-\delta_{m,0}\right)e^{imp_a}\hs,\\
d(m,0,p_a)&=&\left(1-\frac{2m}{M}\right)e^{imp_a}\hs.\nonumber
\ENA
Then for $m=0$ one reproduces the obvious answer
\EQ
\la\sm_+\sm_-\ra_{_N}=\frac{1}{2}+\frac{1}{2}\la\sm_z\ra_{_N}=1-\frac{N}{M}
\hs.
\EN
For $m>0$ the equal-time mean value is represented as follows:
\EQA
\la\snd_+\snu_-\ra_{_N}&=&\deriv{z}\left.\adet_{_N}\left[s+zr^{(+)}\right]
\right\vert_{z=0}=\nonumber\\
&=&\adet_{_N}\left[s+r^{(+)}\right]-\adet_{_N}\left[s\right]\hs,
\hspace{1.5cm} m>0\hs,\label{sofive}
\ENA
where matrix elements of $N\times N$ matrices $s$ and
$r^{(+)}$ are
\EQA
s_{ab}&=&\delta_{ab}\left(1-\frac{2m}{M}\right)-\frac{2}{M}
\left(1-\delta_{ab}\right)\frac{\sin\frac{m}{2}(p_a-p_b)}{\tan\frac{1}{2}
(p_a-p_b)}\hs,\\
r^{(+)}_{ab}&=&\frac{1}{M}e^{\frac{im}{2}(p_a+p_b)}\hs,
\label{soseven}
\ENA

Analogous representations are valid also for mean value
$\la\snd_-(t_2)\snu_+(t_1) \ra_{_N}$; as
\EQ
\la\snu_-(t_1)\snd_+(t_2) \ra_{_N}=\la\snd_-(t_2)\snu_+(t_1)
\ra_{_N}^{\ast}
\hs,
\EN
it is again considered in region $m\geq 0$, $-\infty<t\equiv(t_2-t_1)

<+\infty$. In the time-dependent case, one has
\EQA
\la\snd_-(t_2)\snu_+(t_1) \ra_{_N}&=&e^{2iht}\deriv{z}\left.\adet_{_N}
\left[S+zR^{(-)}\right]\right\vert_{z=0}=\nonumber\\
&=&e^{2iht}\left\{\adet_{_N}\left[S+R^{(-)}\right]-\adet_{_N}\left[S\right]
\right\}\hs,\spz m\geq 0\hs,\label{soeight}
\ENA
where $N\times N$ matrix $S$ is just the same as in (\ref{sthree}),
(\ref{sfour})
and the matrix elements of $N\times N$ matrix $R^{(-)}$ are given as
\EQ
R^{(-)}_{ab}=\frac{1}{M}e_-(m,t,p_a) e_-(m,t,p_b)\hs,
\EN
function $e_-$ being defined in (\ref{ssix}).
In the equal-time case, the obvious relations
\EQ
\la\sm_-\sm_+\ra_{_N}=\frac{1}{2}-\frac{1}{2}\la \sm_z\ra_{_N}=\frac{N}{M}
\EN
and
\EQ
\la\snd_-\snu_+\ra_{_N}=\la \snu_+\snd_-\ra_{_N}=\la\snd_+\snu_-
\ra_{_N}^{\ast} \hspace{1.5cm}(n_2 > n_1)
\EN
are reproduced.

So the determinant representations for time-dependent (and equal-time)
normalized mean values of products of operators
$\s_+$, $\s_-$ on the finite lattice are given.

To conclude this Section let us discuss the relation of the determinant
representation (\ref{sofive}) for the equal-time  normalized mean value of
operators
$\s_+$, $\s_-$ to representation  (\ref{gener}) for the normalized mean
value of operator $\expaq$. Due to formulae (\ref{fermion}) transforming
the
XX0 model to the free fermion model on a lattice, it is quite natural to
expect
that equal-time correlator (\ref{sofive}) could be
expressed in terms of the first minors of matrix ${\cal M}(m-1)$ entering
representation (\ref{gener}),  with parameter $\alpha$ set equal to
$i\pi$. In fact, as shown in the next Section, due to the possibility of
introducing an arbitrary parameter $c$ into the determinant representation
of the form factors (see (\ref{ftone})),  we can also rewrite
representation
(\ref{sofive}) as
\EQ
\la \snd_+\snu_-\ra_{_N}=\deriv{z}\left.\adet_{_N}\left[{\tilde s}+
z r^{(+)} \right]\right\vert_{z=0}\hs,\hspace{2cm}
m\equiv n_2-n_1 > 0\hs, \label{sttwo}
\EN
with the same matrix $r^{(+)}$ (\ref{soseven}) and
\EQ
{\tilde s}_{ab}=s_{ab}+\frac{ic_1}{M}
e^{-\frac{im}{2}(p_a-p_b)}+\frac{ic_2}{M}e^{\frac{im}{2}(p_a-p_b)}
\EN
(the r.h.s. of (\ref{sttwo}) does not depend on arbitrary complex
parameters
$c_1$, $c_2$). Chosing,  $e.g.$, $c_1=c_2=-i$ one gets
\EQ
{\tilde s}_{ab}=\delta_{ab}\left[1-\frac{2(m-1)}{M}\right]-\frac{2}{M}
\left(1-\delta_{ab}\right)\frac{\sin\frac{m-1}{2}(p_a-p_b)}{\sin\frac{1}{2}
(p_a-p_b)}\hs,\label{stfour}
\EN
which are exactly the matrix elements of matrix ${\cal M}(m-1)$ appearing
in (\ref{gener}) after setting $\alpha=i\pi$.

\resection{Derivation of Representations for Time Dependent Mean Values of
$\s_+$, $\s_-$ on the Finite Lattice}

In this Section representations (\ref{sthree})
and (\ref{soeight}) for the normalized mean values of operators $\s_+\s_-$
and $\s_-\s_+$ are proved. We begin with the normalized mean value
(\ref{sone}) inserting the complete set of states
$\mid\Psi_{N+1}(\{q\})\ra$ between operators $\snd_+(t_2)$ and
$\snu_-(t_1)$. Taking into account the normalization (\ref{norm})
of eigenstates and representation (\ref{fsix}) for form factors
(\ref{defform})-(\ref{ffour}), we readily get
\EQA
\la\snd_+(t_2)\snu_-(t_1)\ra_{_N}&=&\frac{1}{M^{2N+1}}\sum_{\{q\}}
\exp\left[ im(\sum_{1}^{N+1} q_a - \sum_1^N p_b)-it(\sum_1^{N+1}\eps(q_a)
-\sum_1^N\eps(q_b))\right]\cdot\nonumber\\
&&\hspace{4cm} \cdot {\cal F}^2_{_N}(\{q\},\{p\})\hs.\label{seone}
\ENA
Here $\eps(p)=-4\cos p+2h$ is the energy of quasiparticles and the sum is
taken over all the different sets of momenta $\{q\}$ = $q_1,...,q_{N+1}$
of intermediate states (similarly to the summation in (\ref{five})).
As the expression under the sum is symmetric under
permutations of $q$'s (being equal to zero whenever two of
the $q$'s coincide), one can change the sum in (\ref{seone}) for the sum
over all the  permitted values of each $q_a$:
\EQ
\sum_{\{q\}}  \hs\longrightarrow\hs\frac{1}{(N+1)!}\sum_{q_1}\hs
.. \hs\sum_{q_{N+1}}\hs,\label{sumsum}
\EN
where the sum over $q_a$'s ($a=1, ... , N+1$) is to be understood as in
(\ref{soone}):
\EQ
\sum_{q_a}f(q_a)=\sum_{j=1}^M f\left((q_a)_{_j} \right)
\EN
and permitted values $(q_a)_{_j}$ (the same for each $q_a$) are given by
(\ref{sotwo}).

Let us now use representation (\ref{fothree})
\EQN
{\cal F}_{_N}(\{q\},\{p\})=\sum_{Q} (-1)^{[Q]}\prod_{a=1}^N\cot\frac{1}{2}
(q_{Q_a}-p_a)\hs,
\ENN
for one of the two form factors ${\cal F}_{_N}$ in (\ref{seone}). As
in (\ref{seone}) there is the  sum over all the $q_a$'s, and form factor
${\cal F}_{_N}(\{q\},\{p\})$ is antisymmetrical under permutations of
$q_a$'s ($a=1,...,N+1$), the sum over permutations $Q$: ($1,...,N+1$)
$\rightarrow$ ($Q_1,...,Q_{N+1}$) can be written as
\EQN
\sum_Q(-1)^{[Q]}\prod_{a=1}^N\cot\frac{1}{2}(q_{Q_a}-p_a)\hs
\longrightarrow\hs(N+1)!\prod_{a=1}^N\cot\frac{1}{2}(q_a-p_a)
\ENN
without changing the l.h.s. of (\ref{seone}).

Now using representation (\ref{fseven})
\EQN
{\cal F}_{_N}(\{q\},\{p\})=\left[1+\deriv{z} \right]\left.\adet_{_N} A(z)
\right\vert_{z=0}
\ENN
for the remaining factor ${\cal F}_{_N}$ and taking into account that
$\adet_{_N} A(z)$ is a linear function of $z$, one rewrites (\ref{seone})
as
\EQA
\la\snd_+(t_2)\snu_-(t_1)\ra_{_N}&=&\exp\left[-2iht-4it\sum_{b=1}^N
\cos p_b-im\sum_{b=1}^N p_b\right]\cdot\nonumber\\
&&\hs\cdot\sum_{q_1}...\sum_{q_{N+1}}\left\{\frac{1}{M}\exp\left[
4it\cos q_{N+1}+imq_{N+1}\right]+\deriv{z}\right\}\cdot\nonumber\\
&&\hs\left.\cdot\adet_{_N}{\cal U}(z)\right\vert_{z=0}\hs,
\label{setwo}
\ENA
where the $N\times N$ matrix ${\cal U}(z)$ is given as
\EQA
{\cal U}(z)&=&{\cal U}^{(1)}-\frac{z}{M}{\cal U}^{(2)}\hs,\nonumber\\
{\cal U}^{(1)}_{ab}&=&\frac{1}{M^2}\exp\left[4it\cos q_a+imq_a\right]
\hs\cot\frac{1}{2}(q_a-p_a)\hs\cot\frac{1}{2}(q_a-p_b)\hs,\nonumber\\
{\cal U}^{(2)}_{ab}&=&\frac{1}{M^2}\exp\left[4it\cos q_a+imq_a\right]\hs
\exp\left[4it\cos q_{N+1}+imq_{N+1}\right]\cdot\nonumber\\
&&\hs\cdot\cot\frac{1}{2}(q_a-p_a)\hs\cot\frac{1}{2}(q_{_{N+1}}-p_b)\hs.
\label{sethree}
\ENA
It should be noted that the rank of matrix ${\cal U}^{(2)}$ is equal to
one (all its rows are proportional to each other); hence $\adet_{_N}
{\cal U}(z)$
is a linear function of $z$.

Let us consider now summation in $q_{_{N+1}}$ in (\ref{setwo}). As
$\adet_{_N} {\cal U}(z=0)$ does not contain $q_{_{N+1}}$ (all the
dependence on $q_{_{N+1}}$ is in matrix ${\cal U}^{(2)}$), one has
\EQ
\left. \frac{1}{M}\sum_{q_{_{N+1}}}\exp\left[4it\cos
q_{_{N+1}}+imq_{_{N+1}}
\right] \adet_{_N} {\cal U}(z) \right\vert_{z=0}=\left. g(m,t)
\adet_{_N}{\cal U}(z)
\right\vert_{z=0}\hs,\label{sefour}
\EN
with function $g(m,t)$ defined in (\ref{sseven}). On the other hand, using
the fact that the rank of matrix ${\cal U}^{(2)}$ is equal to one,
we conclude that
\EQ
\sum_{q_{N+1}}\deriv{z}\adet_{_N}{\cal U}(z)=\deriv{z}\adet_{_N}\left[
{\cal U}^{(1)}-\frac{z}{M}{\tilde{\cal U}}^{(2)} \right]\hs, \label{sesix}
\EN
where matrix ${\cal U}^{(1)}$ is just the same as in (\ref{sethree}) and
\EQ
{\tilde{\cal U}}^{(2)}_{ab}=\frac{1}{M}\exp\left[4it\cos q_a+im q_a
\right]\hs\cot\frac{1}{2}(q_a-p_a)\hs e(m,t,p_b)
\label{seseven}
\EN
with function $e(m,t,p_b)$ defined in (\ref{seight}). So the summation
over $q_{_{N+1}}$ can be done ``inside'' the determinant.

The summations over the remaining $q_a$'s ($a=1, ... ,N$) can be also
reduced
to the summations
of matrix elements of ${\cal U}(z)$, because momentum $q_a$ enters
only the $a^{th}$ row  of this matrix. Therefore one comes to the following
expression:
\EQA
\la\snd_+(t_2)\snu_-(t_1)\ra_{_N}&=&\exp\left[-2iht-4it\sum_{b=1}^N
\cos p_b-im\sum_{b=1}^N p_b \right]\cdot\nonumber\\
&&\hs\cdot\left[g(m,t)+\deriv{z}\right]\left.\adet_{_N}\left[
{\tilde{\cal U}}^{(1)}-z R\right]\right\vert_{z=0}\hs,\label{senine}
\ENA
with
\EQA
{\tilde{\cal U}}_{ab}^{(1)}&=&\frac{1}{M^2}\sum_q\exppq\hs\cot\frac{1}{2}
(q-p_a)\hs\cot\frac{1}{2}(q-p_b)\hs,\label{seten}\\
R_{ab}&=&\frac{1}{M}e(m,t,p_a)e(m,t,p_b)\hs.\label{seoone}
\ENA
The diagonal elements of matrix ${\tilde{\cal U}}^{(1)}$ can be written
as
\EQ
{\tilde{\cal U}}^{(1)}_{aa}=d(m,t,p_a)-\frac{1}{M}g(m,t)\hs,\label{seotwo}
\EN
with function $d(m,t,p_a)$ defined in (\ref{snine}).
Taking into account the identity
\EQN
\cot\frac{1}{2}(q-p_a)\hs\cot\frac{1}{2}(q-p_b)=\cot\frac{1}{2}(p_a-p_b)
\hs\left[\cot\frac{1}{2}(q-p_a)-\cot\frac{1}{2}(q-p_b)\right]-1\hs,
\ENN
off-diagonal elements of matrix ${\tilde{\cal U}}^{(1)}$ turn out to be:
\EQ
{\tilde{\cal U}}^{(1)}_{ab}=\frac{1}{M}\frac{1}{\tan\frac{1}{2}(p_a-p_b)}
\left[e(m,t,p_a)-e(m,t,p_b) \right]-\frac{1}{M}g(m,t)\hs.\label{seofour}
\EN
Let us now ``insert'' the exponential factor in the r.h.s. of
(\ref{senine})
inside the determinant; this means to multiply the $a^{th}$ row by
\EQN
\exp\left[ -\frac{im}{2}p_a-2it\cos p_a\right]\hs,
\ENN
and the $b^{th}$ column by
\EQN
\exp\left[ -\frac{im}{2}p_b-2it\cos p_b\right]\hs.
\ENN
Taking into account definition (\ref{ssix}) of functions $e_+$, $e_-$, one
comes just to representation (\ref{sthree}), which is thus proved.

It is to be mentioned that instead of representations (\ref{fseven}) and
(\ref{fothree}) for the
form factors in (\ref{seone}), one can also use representation
(\ref{ftone})
\EQN
{\cal F}_{_N}(\{q\},\{p\})=\left[ 1+\deriv{z}\right]\left.\adet_{_N}
A(z,c_1)
\right\vert_{z=0}
\ENN
for one of them, and the form analogous to (\ref{fothree})
\EQN
{\cal F}_{_N}(\{q\},\{p\})=\sum_{Q}(-1)^{[Q]}\prod_{a=1}^N\left[\cot
\frac{1}{2} (q_{Q_a}-p_a) + c_2 \right]
\ENN
for the other one; here $c_1$, $c_2$ are arbitrary complex constants; the
answer, of course, does not depend on $c_1$, $c_2$. This alternative
procedure
gives, instead of (\ref{sthree}),  the following representation
\EQA
\la\snd_+(t_2)\snu_-(t_1)\ra_{_N}&=&e^{-2iht}\left[g(m,t)+\deriv{z} \right]
\cdot\nonumber\\
&&\spz\cdot  \left. \mbox{det}_N\left[{\tilde S}(c_1,c_2)
-z{\tilde R}^{(+)}
(c_1,c_2)\right]\right\vert_{z=0}\hs,\label{seofive}
\ENA
where matrix elements of $N\times N$ matrices ${\tilde S}$, ${\tilde
R}^{(+)}$
are
\EQA
{\tilde S}_{ab}(c_1,c_2)&=&S_{ab}+\frac{c_1}{M}e_+(m,t,p_a)e_-(m,t,p_b)
+\frac{c_2}{M}e_-(m,t,p_a) e_+(m,t,p_b)+\nonumber\\
&&\spz+\frac{c_1 c_2}{M}
g(m,t)e_-(m,t,p_a) e_-(m,t,p_b)\hs;\nonumber\\
{\tilde R}^{(+)}_{ab}(c_1,c_2)&=&\frac{1}{M}\left[e_+(m,t,p_a)+c_2
e_-(m,t,p_a) g(m,t) \right]\cdot\nonumber\\
&&\hs\cdot\left[ e_+(m,t,p_b)+c_1 e_-(m,t,p_b) g(m,t) \right]\cdot\nonumber
\ENA
where $S_{ab}$ = ${\tilde S}_{ab}(0,0)$ is given by (\ref{sfour}).
For $t=t_2-t_1=0$ and $m=n_2-n_1>0$ (since then $g(m,t=0)=0$) one
gets just representation (\ref{sttwo}) for the equal-time correlator.

Let us now consider the other normalized mean value:
\EQA
\la\snd_-(t_2)\snu_+(t_1)\ra_{_N}&=&\frac{\la\psin(\{p\})\mid\snd_-(t_2)
\snu_+(t_1)\mid\psin(\{p\})\ra}{\la\psin(\{p\})\mid\psin(\{p\})\ra}\hs,
\nonumber\\
m\equiv n_2-n_1\geq 0;&&\hspace{1cm}-\infty<t\equiv t_1-t_2<+\infty.
\label{cormp}
\ENA
Inserting the complete set of $N-1$-particle states
$\mid\Psi_{n-1}(\{q\})\ra$ ($\{q\}=q_1, ... ,q_{_{N-1}}$)
and using representation (\ref{fsix}) for  the appearing form factors,
one gets
\EQA
\la\snd_-(t_2)\snu_+(t_1)\ra_{_N}&=&\frac{1}{M^{2N-1}}\sum_{\{q\}}
\exp\left[-im(\sum_{1}^{N} p_a - \sum_1^{N-1} q_b)+it(\sum_1^{N}\eps(p_a)
-\sum_1^{N-1}\eps(q_b))\right]\cdot\nonumber\\
&&\hspace{4cm} \cdot {\cal F}^2_{_{N-1}}(\{p\},\{q\})\hs.\label{seonebis}
\ENA
It is to be emphasized that now the number of external momenta $p_a$
($a=1, ... ,N$) is larger than the number of intermediate (summed over)
momenta $q_b$ ($b=1, ... , N-1$); hence the order of arguments in
${\cal F}_{_N}(\{p\},\{q\})$. The sum over different sets $\{q\}$ in
(\ref{seonebis}) can be changed (analogously to (\ref{sumsum})) to
independent
sums over all the permitted values of each $q_a$:
\EQ
\sum_{\{q\}}\hs\longrightarrow\hs\frac{1}{(N-1)!}\sum_{q_1} ...
\sum_{q_{N-1}}
\hs.
\EN
Representing form factor ${\cal F}_{_{N-1}}(\{p\},\{q\})$ by means of
(\ref{fothree}),
\EQ
{\cal F}_{_{N-1}}(\{p\},\{q\})=\sum_{Q} (-1)^{[Q]}\prod_{a=1}^{N-1}\cot
\frac{1}{2}(p_{Q_a}-q_a)\hs,
\EN
(where the sum is now over permutations $Q$: ($1, ... , N$) $\goto$
($Q_1, ... ,Q_N$)) one represents (\ref{seonebis}) as
\EQA
\la\snd_-(t_2)\snu_+(t_1)\ra_{_N}&=&\frac{1}{M^{2N-1}(N-1)!}
\exp\left[2iht- im \sum_1^N p_a-4it\sum_1^N\cos
(p_a)\right]\cdot\nonumber\\
&&\hspace{1.5cm} \cdot \sum_{P,Q}(-1)^{[P]+[Q]}\prod_{a=1}^{N-1}
 {\cal S}(p_{Q_a},p_{P_a},m,t)\hs.\label{ciao}
\ENA
Here function ${\cal S}$ is introduced,
\EQA
{\cal S}(p_a,p_b,m,t)&\equiv&\sum_q\exp\left[imq+4it\cos q \right]\cot
\frac{1}{2}(q-p_a)\cot\frac{1}{2}(q-p_b)\hs;\nonumber\\
{\cal S}(p_a,p_a,m,t)&=&M^2 d(m,t,p_a)-M g(m,t)\hs,\label{ciaodue}\\
{\cal S}(p_a,p_b,m,t)&=&\frac{M}{\tan\frac{1}{2}(p_a-p_b)}\left[e(m,t,p_a)-
e(m,t,p_b) \right]-M g(m,t)\hs,\spz {\rm if} \hs a\not=b\hs,\nonumber
\ENA
with functions $g$, $e$, $d$ defined in (\ref{sseven})-(\ref{snine}).
The sum over $P$, $Q$ in (\ref{ciao}) is the sum over all the permutations,
$P$: ($1, ... , N$) $\goto$ ($P_1, ... , P_N$) and
$Q$: ($1, ... , N$) $\goto$ ($Q_1, ... , Q_N$). It is not difficult
to notice that this sum is just the sum of all the first minors (multiplied
by $(N-1)!$) of $N\times N$ matrix ${\tilde{\cal S}}$
with matrix elements
\EQ
{\tilde {\cal S}}_{ab}\equiv{\cal S}(p_a, p_b, m,t)\hs,
\EN
and can therefore be represented in the form
\EQ
\sum_{P,Q} (-1)^{[P]+[Q]}\prod_{a=1}^{N-1} {\cal S}(p_{Q_a},p_{Q_b},m,t)=
(N-1)!\left.\deriv{z}\adet_{_N}\left[{\tilde{\cal S}}+{\tilde z}
\right]\right\vert_{z=0}\hs.\label{ciaotre}
\EN
Here all the elements of matrix ${\tilde z}$ are the same:
\EQ
{\tilde z}_{ab}= z\hs,
\EN
($z$ is as usual a complex parameter) so that in fact $\adet_{_N}\left[
{\tilde{\cal S}} +{\tilde  z} \right]$ is a linear function of $z$.

Representing now the sums over permutations in (\ref{ciao}) by means
of (\ref{ciaotre}) one  easily transform the representation obtained
 just to the form (\ref{soeight}) which is hence proved.

\resection{Correlators in the Thermodynamic Limit; the Case of Zero
Temperature}

The most interesting from the physical point of view are correlators of the
model in the
thermodynamical limit where the total number $M$ of sites of the
lattice goes to infinity, $M\goto\infty$. In this Section the
thermodynamical
limit at zero temperature is discussed, correlators being defined as
the normalized mean values of corresponding operators with respect
to the ground state of the model (see (\ref{zerotemp})).

The ground state $\mid\Omega_M\rangle$  for fixed magnetic field
$h$ ($0\leq h<h_c\equiv 2$, see discussion in Section 2) is obtained by
filling with quasiparticles all the permitted vacancies with momenta
$p_a$ inside the Fermi zone, $-k_F\leq p_a\leq k_F$, where $k_F$
is the Fermi momentum  (\ref{fermi}) $k_F$ $=$ $\arccos(h/2)$.
In the thermodynamical limit the permitted values of momenta corresponding
to the ground state fill the whole interval $[-k_F,k_F]$,
with the number of quasiparticles in the ground state, $N=M k_{_F}/\pi$,
see (\ref{allowed}), going to infinity. However, density $D=\frac{N}{M}=
\frac{k_F}{\pi}$ of quasiparticles  in the ground state remains finite,
the magnetization $\langle\s_z\rangle$ being (see (\ref{magnet})):
\EQ
\langle\s_z \rangle=\langle\sm_z \rangle=1-2D=1-\frac{2 k_F}{\pi}\hs.
\label{etwo}
\EN

Let us consider first the thermodynamical limit of generating functional
(\ref{mean1})
\EQ
\la\expaq\ra=\frac{\la\Omega\mid\expaq\mid\Omega\ra}{\la\Omega\mid
\Omega\ra}\hs,
\EN
where $\mid\Omega\ra$ is the ground state in the limit: $\mid\Omega\ra
=\lim_{M\goto\infty}\mid\Omega_M\ra$, magnetic field $h$ being fixed.
Taking the corresponding limit in formula (\ref{gener}) one comes
to the following representation for the generating functional:
\EQ
\la\expaq\ra=\left.\adet (\ic+\gamma\mc)\right\vert_{\gamma=e^{\alpha}-1}
\label{fgi}
\EN
In the r.h.s. there is the Fredholm determinant
of linear integral operator $\mc(m)$ acting on functions $f(p)$ on interval
$[-k_F,k_F]$ according to the rule:
\EQ
(\mc f)(p)=\frac{1}{2\pi}\intk dq \hs M(p,q) f(q)
\EN
with kernel $M(p,q)$ given as
\EQ
M(p,q)=\frac{\sin\frac{m}{2}(p-q)}{\sin\frac{1}{2}(p-q)}\hs,\label{esix}
\EN
Operator $\ic$ in (\ref{fgi}) is the unit operator, with kernel
$2\pi\delta(p-q)$.

To derive this representation one has to notice that $N\times N$ matrix
${\cal M}$ in (\ref{gener}) acts on arbitrary N-component vector $f_a$
($a=1,...,N$) as follows
\EQA
({\cal M}(m) f)_a&=&f_a+\frac{\gamma m}{M}f_a+\frac{\gamma}{M}
\sum_{\begin{array}{c}\\ [-12mm] _{b=1}\\ [-5mm]
_{b\not=a}\end{array}}^N
\frac{\sin \frac{m}{2}(p_a-p_b)}{\sin\frac{1}{2}(p_a-p_b)}
f_b\hs,\nonumber\\
&&\spz \gamma\equiv e^{\alpha}-1\hs.\label{eightseven}
\ENA
In the thermodynamical limit, due to (\ref{allowed}), one should replace
the
sum by the integral
\EQ
\frac{1}{M}\sum_{a=1}^{N}\hs\longrightarrow\hs\frac{1}{2\pi}\intk dq
\label{eeight}
\EN
coming just to (\ref{fgi}) (the second term in the r.h.s. of
(\ref{eightseven})
is included naturally to operator $\mc(m)$).
Let us also mention that for $\alpha=-\infty$ ($\gamma=-1$) the generating
functional possesses a clear physical meaning giving the probability
$P(m)$ for all spins in an interval of length $m$ to be up in the ground
state
($i.e.$ that there are no quasiparticles on this interval):
\EQ
P(m)=\la\expaq\ra_{\alpha=-\infty}=\adet(\ic-\mc)\hs.
\EN

Let us consider now the time-dependent correlator
\EQ
g^{(0)}_{zz}(m,t)\equiv\langle\s^{(n_2)}_z
(t_2)\s^{(n_1)}_z(t_1)\rangle\equiv
\frac{\langle\Omega
\mid\s^{(n_2)}_z (t_2)\s^{(n_1)}_z(t_1)\mid\Omega\rangle}{\langle
\Omega\mid\Omega\rangle}\hs,
\EN
in the thermodynamical limit. Replacing sums over $a$ ($a=1,...,N$) in
formula
(\ref{corrzt}) with integrals over the Fermi zone, as in (\ref{eeight}),
and
sum over all the particles $j$ ($j=1,...,M$) as
\EQ
\sum_{j=1}^M \hs\longrightarrow\hs\frac{1}{2\pi}\intp dq\hs,
\EN
one gets
\EQA
g^{(0)}_{zz}(m,t)&=&\la\s_z\ra^2-
\frac{1}{\pi^2}\left\vert\intk dp\hs \exp\left[imp+4it\cos p\right]
\right\vert^2+\label{eotwo}\\
&&\hs+\frac{1}{\pi^2}\intk dp\hs\exp\left[-imp-4it\cos p \right]\intp dq\hs
\exp\left[imq+4it\cos q \right]\hs,\nonumber
\ENA
where $m=m_2-m_1$, $t=t_2-t_1$, and $\la\s_z\ra$ is the magnetization
(\ref{etwo}); explicit expression (\ref{energy}) for the energy,
$\eps(p)=-4\cos p +2h$,
has been taken into account. This last result has already been obtained
from
a different procedure in \cite{two}. In the equal-time case ($t=0$), the
last integral
is equal to zero for $m\not=0$, while the other ones can be calculated
explicitly, reproducing the well known answer
\EQ
\la\s^{(n_2)}_z\s_z^{(n_1)}\ra=\la\s_z\ra^2-\frac{4}{\pi^2}
\frac{\sin^2 mk_F}{m^2}\hs,\hspace{2cm} m\equiv n_2-n_1\not=0\hs.
\EN

Let us turn now to correlators of operators
$\s_+$, $\s_-$, considering first correlator (\ref{corrp})
\EQ
g^{(0)}_+(m,t)\equiv\la\s^{(n_2)}_+(t_2)\s_-^{(n_1)}(t_1)\ra\hs,\label{eofou
r}
\EN
where as usual $m=n_2-n_1$, $t=t_2-t_1$. Due to the relations
\EQ
g^{(0)}_+(m,t)=g^{(0)}_+(-m,t)=\left[ g^{(0)}_+(-m,-t)\right]^{\ast}\hs,
\EN
it is sufficient to consider the region
\EQ
m\geq 0\hs,\hspace{2cm} t\geq 0\hs.
\EN
The determinant representation for this correlator is obtained by performing
the thermodynamical limit in representation (\ref{sthree}) for finite
number $M$ of sites; the
ground state of the model at finite $M$ should be taken as state
$\mid\Psi_N(\{p\})\ra$ with respect to which the mean value is taken in
eq. (\ref{sone}), (\ref{sthree}).

As shown in Appendix, functions $g$, $e$, $d$
(\ref{sseven})-(\ref{snine}) entering the
representation for the finite lattice should be changed to functions
$\gc$, $\ec$, $\dc$, respectively, in the thermodynamical limit:
\EQA
\gc(m,t)&=&\frac{1}{2\pi}\intp dq\hs\exp\left[imq+4it\cos q\right]=
\nonumber\\
&=&i^m J_m(4t)\hs,\label{ounosette}
\ENA
($J_m$ is a Bessel function)
\EQA
\ec(m,t,p)&=&\frac{1}{2\pi}\intp dq\hs\frac{\exppq-\exppp}{\tan\frac{1}{2}
(q-p)}\hs\equiv\nonumber\\
&\equiv&\frac{1}{2\pi}{\cal P}\intp dq \hs\frac{\exppq}{\tan\frac{1}{2}
(q-p)}\hs,
\ENA
and
\EQ
\dc(m,t,p)=\exppp+\frac{2}{M}\frac{\partial\hs}{\partial p}\ec(m,t,p)\hs.
\label{eonine}
\EN
Then, in analogy with the derivation of formula (\ref{fgi}) for
generating functional $\la\expaq\ra$, one obtains the representation
for correlator $g^{(0)}_+(m,t)$ (\ref{eofour}) in
the thermodynamical limit in terms of the Fredholm determinant
of a linear integral operator:
\EQA
g^{(0)}_+(m,t)&=&\left. e^{-2iht}\left[\gc(m,t)+\frac{\partial\hs}{\partial
z} \right]\adet\left[\ic+\vc-z\rc^{(+)} \right]\right\vert_{z=0}\hs=
\nonumber\\
&=&e^{-2iht}\left\{\left[\gc(m,t)-1\right]\adet\left[\ic+\vc\right]+
\adet\left[\ic+\vc-\rc^{(+)}\right]\right\}\hs,\label{etwenty}
\ENA
where $\ic$ is again the identity operator, and linear operators
$\vc$, $\rc^{(+)}$ acting  on functions $f(p)$ on segment
$[-k_F,k_F]$
\EQA
\left(\vc f\right)(p)&=&\frac{1}{2\pi}\intk dq\hs V(p,q) f(q)\hs,
\nonumber\\
\left(\rc^{(+)}f\right)(p)&=&\frac{1}{2\pi}\intk dq\hs R^{(+)}(p,q)f(q)\hs,
\ENA
possess kernels
\begin{eqnarray}
V(p,q) &=& \frac{\ec_+(p) \ec_-(q)-\ec_-(p)
\ec_+(q)}{\tan\frac{1}{2}(p-q)}-
\gc(m,t) \ec_-(p) \ec_-(q)\hs,\label{ettwo} \\
R^{(+)}(p,q) &=& \ec_+(p) \ec_+(q)\hs.\label{etthree}
\end{eqnarray}
Functions $\ec_+$, $\ec_-$, are defined analogously to $e_+$, $e_-$ in
(\ref{ssix}):
\begin{eqnarray}
\ec_-(p) &\equiv& \ec_-(m,t,p) = \exp[-\frac{i}{2}mp-2it\cos
p]\hs,\nonumber \\
\ec_+(p) &\equiv& \ec_+(m,t,p) = \ec_-(p) \ec(m,t,p) \hs.\label{etfour}
\end{eqnarray}
It is to mention that the second term in the right hand side of expression
(\ref{eonine}) for function $\dc$ is included naturally to operator $\vc$.

Thus the representation for correlator $g^{(0)}_+(m,t)$ (\ref{eofour})
is obtained.

Let us  consider now correlator (\ref{corrm})
\EQ
g^{(0)}_-(m,t)\equiv\la \s^{(n_2)}_-(t_2)\s^{(n_1)}_+(t_1)\ra\hs.
\EN
Again, due to the properties
\EQ
g^{(0)}_-(m,t)=g^{(0)}_-(-m,t)=\left[g^{(0)}_-(-m,-t) \right]^{\ast}\hs,
\EN
we consider it only in region $m\geq 0$, $t\geq 0$. Using
expression (\ref{soeight}) at finite $M$, one derives the following
representation in the thermodynamical limit:
\EQA
g^{(0)}_-(m,t)&=&e^{2iht}\frac{\partial\hs}{\partial z} \left.\adet
\left[\ic+\vc+z\rc^{(-)} \right]\right\vert_{z=0}=\nonumber\\
&=&e^{2iht}\left\{\adet\left[\ic+\vc+\rc^{(-)} \right]-\adet
\left[\ic+\vc\right]\right\}\hs,\label{etseven}
\ENA
where $\vc$ is just the same linear operator as in representation
(\ref{etwenty}) for correlator $g^{(0)}_+$, see (\ref{ettwo}), and
the kernel of operator $\rc^{(-)}$ is
\EQ
R^{(-)}(p,q)=\ec_-(m,t,p)\hs \ec_-(m,t,q)\hs.\label{eteight}
\EN

In the equal-time case ($t=0$) functions $\gc$, $\ec$ are calculated
explicitly to be
\EQA
\gc(m,0)&=&\delta_{m,0}\hs,\nonumber\\
\ec(m,0)&=&i\left[1-\delta_{m,0}\right]e^{imp}\hs,
\ENA
so that for equal time correlator $g^{(0)}_+(m,0)$ one gets
the representation (which is, of course, the thermodynamical limit
of representation (\ref{sofive})):
\EQA
g^{(0)}_+(m)&=&\la \s_+^{(n_2)}\s_-^{(n_1)}\ra=\nonumber\\
&=&\frac{\partial\hs}{\partial z}\adet\left[\ic+\vcs+z\rcs^{(+)}\right]=
\nonumber\\
&=&\adet\left[\ic+\vcs+\rcs^{(+)}\right]-\adet\left[\ic+\vcs\right]\hs,
\label{blabla}\ENA
where the kernels of operator $\vcs$, $\rcs^{(+)}$ (acting on interval

$[-k_F, k_F]$) are
\EQA
v(p,q)&=&-2\frac{\sin\frac{m}{2}(p-q)}{\tan\frac{1}{2}(p-q)}\hs,\\
r^{(+)}(p,q)&=&\exp\left[\frac{im}{2}(p+q) \right]\hs.\label{ethtwo}
\ENA

This answer can also be put into the following form  (see (\ref{stfour})
and
the discussion at the end of Section 6):
\EQ
g^{(0)}_+(m)=\deriv{z}\adet\left[\ic+{\hat w}+z\rcs^{(+)}\right]\hs,\spz
m>0\hs,
\EN
where operator $\rcs^{(+)}$ is the same as in (\ref{ethtwo}) and the kernel
of operator ${\hat w}$ is
\EQ
W_0(p,q)=-2\frac{\sin\frac{m-1}{2}(p-q)}{\sin\frac{1}{2}(p-q)}\hs,
\label{ethfour}
\EN
so that in fact the representation of correlator $g^{(0)}_+(m)$ involves
the first Fredholm minors of the same linear integral operator as in
representation (\ref{fgi}) of the
generating functional $\la\expaq\ra$ at $\alpha=i\pi$ (compare
(\ref{ethfour}) and (\ref{esix})).

\resection{Correlators in the Thermodynamical Limit at Nonzero Temperature}

In this Section we consider the correlators at non-zero temperature ($T>0$)
in the thermodynamical limit.
Temperature dependent correlation functions are defined in the standard
way (\ref{fintemp}). For some operator ${\cal O}$, the temperature mean
value
$\la{\cal O}\ra_{_T}$ is
\EQ
\la {\cal O}\ra_{_T}=\frac{{\rm Sp}\left[e^{-\frac{H}{T}}{\cal O}
\right]}{{\rm Sp}\left[e^{-\frac{H}{T}} \right]}.
\EN
For integrable systems, calculating mean values of this kind is
rather simple \cite{lecture}. From the practical point of
view, one should  only change
the integration measure in the representations of correlators obtained
in the zero temperature case, namely,
\EQ
\intk dq\hs\longrightarrow\hs\intp dq\hs\vtq\hs.
\EN
Here $\vtq\equiv\vartheta(q,h,T)$ is the Fermi weight (\ref{weight}),
\EQ
\vtq=\frac{1}{1+\exp\left[\frac{\eps(q)}{T}\right]}
=\frac{1}{1+\exp\left[\frac{-4\cos q+2h}{T}\right]}\hs,
\EN
describing the momenta distribution of particles at the thermodynamical
equilibrium. Of course, for the XX0 chain, which is equivalent to the free
fermion model, this procedure is quite obvious.

Taking this into account, it is straightforward to extend the
representations of zero temperature correlators to
finite temperatures.

For the generating functional $\la\expaq \ra_{_T}$ one has (see
(\ref{fgi})) the representation in terms of the Fredholm determinant:
\EQ
\la\expaq\ra_{_T}=\adet\left.\left[ \ic+\gamma\mc_T\right]\right
\vert_{\gamma=e^{\alpha}-1} \label{repnf}
\EN
where now $\mc_T$  is an integral operator acting on functions $f(p)$
on interval $[-\pi,\pi]$:
\EQ
\left(\mc_T f\right)(p)=\frac{1}{2\pi}\intp dq\hs M_T(p,q) f(q)\hs,
\label{ncinque}
\EN
with kernel $M_T(p,q)$ obtained from kernel $M(p,q)$ (\ref{esix})
as
\EQA
M_{_T}(p,q)&=&M(p,q)\hs\vtq=\frac{\sin\frac{m}{2}(p-q)}{\sin\frac{1}{2}(p-q)
}
\hs\vtq\hs,\nonumber\\
&&-\pi\leq p,q\leq\pi\hs.
\ENA
Since the similarity transformation,
\EQ
M_{_T}(p,q)\hs\longrightarrow\hs\sqrt{\vtp}\hs
M_{_T}(p,q)\hs\frac{1}{\sqrt{
\vtq}}\hs,\label{symtran}
\EN
does not change the value of the Fredholm determinant,  one can rewrite
representation (\ref{repnf}) as
\EQ
\la\expaq\ra_{_T}=\adet\left.\left[ \ic+\gamma\mc_{_T}^{(S)}\right]\right
\vert_{\gamma=e^{\alpha}-1}
\EN
where operator $\mc_{_T}^{(S)}$ possesses the symmetrical kernel
\EQ
M_{_T}^{(S)}(p,q)=\sqrt{\vtp}\hs\frac{\sin\frac{m}{2}(p-q)}{\sin
\frac{1}{2}(p-q)}\hs\sqrt{\vtq}\hs.
\EN

Consider now the time-dependent correlator of the third spin components.
Representation (\ref{eotwo}) is rewritten for $T>0$ as \cite{two}
\EQA
&&g^{(T)}_{zz}(m,t)\equiv \nonumber\\
&&\hspace{4mm}\equiv\la\s^{(n_2)}_z(t_2)\s^{(n_1)}_z(t_1) \ra_{_T}=
\la\s_z\ra^2-\frac{1}{\pi^2} \left\vert\intp dp\hs\vtp\exppp\right\vert^2+
\label{nseven}\\
&&\hspace{8mm}+\frac{1}{\pi^2}\left(\intp dp\hs\vtp\expmp\right)
\left(\intp dq\hs\exppq\right)\hs.\nonumber
\ENA
In the equal-time case, the known answer is reproduced:
\EQA
&&g^{(T)}_{zz}(m)\equiv\la\s^{(n_2)}_z\s^{(n_1)}_z \ra_{_T}
= \la\s_z\ra^2_T-\frac{1}{\pi^2}\left\vert
\intp dp \vtp e^{imp}\right\vert^2\hs,\nonumber\\
&&\hspace{3cm} (m\equiv n_2-n_1\not=0)
\label{neight}
\ENA
$i.e.$ the correlator is given by the square modulus of the
Fourier transform of Fermi weight. In (\ref{nseven}), (\ref{neight}),
$\la\s_z\ra_{_T}$ is the magnetization at temperature $T$,
\EQ
\la\s_z\ra_{_T}=1-\frac{1}{\pi}\intp dq\hs\vtq\hs,
\EN
with $\la\s_z\ra_{_T}=0$ at $h=0$.

Let us now consider correlators of operators $\s_+$, $\s_-$. For
correlator (\ref{corrp}) one gets (see (\ref{etwenty}))  the following
representation at $T>0$
\EQA
g_+^{(T)}(m,t)&\equiv&\la\s_+^{(n_2)}(t_2)\s_-^{(n_1)}(t_1)\ra_{_T}=
\nonumber\\
&=&e^{-2iht}\left[\gc(m,t)+\frac{\partial \hs}{\partial z}
\right]\left.\adet\left[\ic+\vc_T-z\rc^{(+)}_T\right]\right\vert_{z=0}=
\label{nten}\\
&=&e^{-2iht}\left\{\left[\gc(mt)-1\right]\adet\left[\ic+\vc_T \right]
+\adet\left[\ic+\vc_T-\rc_T^{(+)} \right]\right\}\hs,\nonumber
\ENA
where function $G(m,t)$ (\ref{ounosette}) is the same as in (\ref{etwenty})
and operators $\vc_T$ and $\rc_T^{(+)}$ (acting on interval $[-\pi,\pi]$,
see (\ref{ncinque})) possess kernels
\EQA
V_{_T}(p,q)&=&V(p,q)\vtq\hs,\nonumber\\
R^{(+)}_{_T}&=& R^{(+)}(p,q)\vtq\hs, \label{noone}
\ENA
with functions $V(p,q)$ and $R^{(+)}(p,q)$ defined (for $-\pi\leq p,q
\leq\pi$) by formulae (\ref{ettwo}), (\ref{etthree}). Making similarity
transform, as in (\ref{symtran}), one rewrites (\ref{noone}) as
\EQA
g_+^{(T)}(m,t)
&=&e^{-2iht}\left[\gc(m,t)+\frac{\partial \hs}{\partial z}
\right]\left.\adet\left[\ic+\vc_{_T}^{(S)}-z\rc^{(+,S)}_{_T}\right]
\right\vert_{z=0}\hs,\\
&&\spz m\geq 0\hs, \nonumber
\ENA
where the kernels of operators are symmetrical,
\EQA
V_{_T}^{(S)}(p,q)&=&\frac{E^T_+(p)E^T_-(q)-E^T_-(p)E^T_+(q)}{\tan
\frac{1}{2}(p-q)}
-\gc(m,t) E^T_-(p) E_-^T(q)\hs,\label{nunosei}\\
R^{(+,S)}_{_T}&=& E_+^T(p) E_+^T(q)\hs,\label{nunosette}
\ENA
and functions $E^T_{\pm}(p)\equiv E_{\pm}^{T}(m,t,p)$  are defined as
\EQ
E^T_{\pm}(p)\equiv\sqrt{\vtp}\hs E_{\pm}(p)\hs\label{nunootto};
\EN
for functions $E_{\pm}(m,t,p)$ see (\ref{etfour}).

For correlator $g_-^{(T)}(m,t)$ (\ref{corrm}) one has (see
(\ref{etseven})):
\EQA
g_-^{(T)}(m,t)&\equiv&\la\s_-^{(n_2)}(t_2)\s_+^{(n_1)}(t_1)\ra_{_T}=
\nonumber\\
&=&e^{2iht}\frac{\partial\hs}{\partial z}\left.\adet\left[ \ic+\vc_{_T}+
z\rc^{(-)}_{_T}\right]\right\vert_{z=0}\hs=\nonumber\\
&=&e^{2iht}\frac{\partial\hs}{\partial z}\left.\adet\left[ \ic+
\vc^{(S)}_{_T}+z\rc^{(-,S)}_{_T}\right]\right\vert_{z=0}\hs
\label{nothree}
\ENA
where $\vc_{_T}$, $\vc_{_T}^{(S)}$ are the same operators
(\ref{noone}), (\ref{nunosei}) and kernels of operators
$\rc_T^{(-)}$, $\rc_T^{(-,S)}$  are
\EQ
R^{(-)}_T(p,q)=R^{(-)}(p,q)\hs\vtq\hs.
\EN
(function $R^{(-)}(p,q)$ was defined in (\ref{eteight})) and
\EQ
R_{_T}^{(-,S)}(p,q)=E_-^T(p)\hs E^T_-(q)
\EN
(for function $E^T_-$ see (\ref{nunootto})).
So the representations for the temperature and time dependent correlators
are given.

In the equal-time case these representations
acquire more explicit form; for correlator
$g_+^{T}(m,t=0)$ one gets
\EQA
g_+^T(m)&\equiv&\la\s^{(n_2)}_+\s^{(n_1)}_-\ra_{_T}=\nonumber\\
&=&\frac{\partial\hs}{\partial z}\left.\adet\left[\ic+\vcs_{_T}+z
\rcs^{(+)}_{_T} \right]\right\vert_{z=0}\hs,\hspace{1.5cm}m>0\hs,
\ENA
with operators $\vcs_{_T}$, $\rcs^{(+)}_{_T}$ acting on interval
$[-\pi,\pi]$, their kernels being given (in the symmetrical form) as
\EQA
\vcs_{_T}(p,q)&=&-2\sqrt{\vtp}\hs\frac{\sin\frac{m}{2}(p-q)}{\tan
\frac{1}{2}(p-q)}\hs\sqrt{\vtq}\hs,\nonumber\\
\rcs^{(+)}_{_T}(p,q)&=& \sqrt{\vtp}\hs\exp\left[\frac{im}{2}(p+q)
\right]\hs\sqrt{\vtq}\hs,
\ENA
which is the generalization for the non-zero temperature case of
representation (\ref{blabla}).

\acknowledgements

One of us (A.G. I.) would like to thank the Department of Physics of the
University of Firenze for hospitality. He is also grateful to Laboratoire
de Physique Th\'eorique in Ecole Normale Sup\'erieure de Lyon, where
part of this work was completed, for invitation on the Louis N\'eel chair.

This work was done in the framework of the NATO S.P. Chaos, Order
and Pattern (CRG 901098).

This work was supported by NSF Grant PHY-9107261

\renewcommand{\theequation}{\Alph{section}.\arabic{equation}}
\appsection{Thermodynamical Limit for Functions $g$, $e$ and $d$.}

Functions $g(m,t)$, $e(m,t,p_a)$ and $d(m,t,p_a)$ on the finite lattice are
defined by formulae (\ref{sseven})-(\ref{snine}) as
\EQA
g(m,t)&\equiv&\frac{1}{M}\sum_{q}\exppq\hs,\label{auno}\\
e(m,t,p_a)&\equiv&\frac{1}{M}\sum_{q}\frac{\exppq}{\tan\frac{1}{2}(q-p_a)}
\hs,\label{adue}\\
d(m,t,p_a)&\equiv&\frac{1}{M^2}\sum_{q}\frac{\exppq}{\sin^2\frac{1}{2}(q-p_a
)}
\hs.\label{atre}
\ENA
The sums are taken over all $M$ different permitted values (\ref{sotwo})
of momentum $q$,
\EQ
\sum_{q} f(q)=\sum_{j=1}^M f(q_j)\hs,
\EN
(see (\ref{soone})). As permitted values of momenta $p_a$ never coincide
with
any $q_j$, there are no singularities  in these finite sums, for finite
$M$.

In the thermodynamical limit momenta $q_j$ fill the
interval $[-\pi,\pi]$. As $q_{j+1}-q_{j}=\frac{2\pi}{M}$, (see
(\ref{sotwo})), one makes the following change:
\EQ
\frac{1}{M}\sum_{q}\hs\longrightarrow\hs\frac{1}{2\pi}\intp dq\hs.
\EN
For function $g(m,t)$ the thermodynamical limit is obtained according to
this rule as
\EQA
g(m,t)\hs\longrightarrow\gc(m,t)&\equiv&\frac{1}{2\pi}\intp dq\hs\exppq=
\nonumber\\
&=&i^m J_m(4t)\hs,
\ENA
where $J_m$ is a Bessel function.

For functions $e$, $d$, however, poles on the integration
contour appear which should be taken into account.

Let us first consider $e(m,t,p_a)$. As
\EQ
\sum_{q}\frac{1}{\tan\frac{1}{2}(q-p_a)}=0\hs,\label{asev}
\EN
(see (\ref{sten}), (\ref{sotwo})), one can write for finite $M$ instead of
(\ref{auno})
\EQ
e(m,t,p_a)=\sum_{q}\frac{\exppq-\exp\left[imp_a+4it\cos p_a\right]}{\tan
\frac{1}{2}(q-p_a)}\hs,
\EN
so that in the thermodynamical limit we get
\EQA
e(m,t,p_a)&\longrightarrow&\ec(m,t,p)=\nonumber\\
&=&\frac{1}{2\pi}\intp dq \frac{\exppq-\exppp}{\tan\frac{1}{2}(q-p)}\hs
\equiv\nonumber\\
&\equiv&\frac{1}{2\pi}{\cal P}\intp dq \frac{\exppq}{\tan\frac{1}{2}(q-p)}
\hs, \label{anine}
\ENA
where ${\cal P}$ means ``Principal Value'', and $p$ is some
value of momentum in the interval $[-\pi,\pi]$.

Analogously, using relation (\ref{asev}) and the following equality
\EQ
\frac{1}{M^2}\sum_q\frac{1}{\sin^2\frac{1}{2}(q-p_a)}=1\hs,
\EN
one writes for $d(m,t,p_a)$ for finite $M$:
\EQA
&&\spz d(m,t,p_a)=\exp\left[imp_a+4it\cos p_a\right]+\label{aeleven}\\
[2mm]
&&+\frac{1}{M^2}\sum_q\frac{\exppq-\exp\left[imp_a+4it\cos p_a\right]
\left[1+(im-4it\sin p_a)\sin(q-p_a) \right]}{\sin^2\frac{1}{2}(q-p_a)}
\nonumber
\ENA

The  subtraction in the numerator is easily seen to be equivalent,
for $q\sim p_a$, to the subtraction of the first two terms in the
expansion of $\exppq$ at $q=p_a$, so that the sum in (\ref{aeleven})
is nonsingular in the thermodynamical limit, and can be represented
as an ordinary integral:
\EQA
&&\spz d(m,t,p_a)\hs\longrightarrow\hs\dc(m,t,p)=\exppp+\label{atwelve}\\
[2mm]
&&+\frac{1}{2\pi M}\intp dq \frac{\exppq-\exppp\left[1+(im-4it\sin
p)sin(q-p)
\right]}{\sin^2\frac{1}{2}(q-p)}\hs.\nonumber
\ENA
Finally, using the identity
\EQA
&&\spz\frac{\partial\hs}{\partial p}\ec(m,t,p)\hs=\hs
\frac{\partial\hs}{\partial p}
\frac{1}{2\pi}{\cal P}\intp dq\frac{\exppq}{\tan\frac{1}{2}(q-p)}=
\nonumber\\ [2mm]
&=&\frac{\partial\hs}{\partial p}\left(\frac{1}{2\pi}\intp dq\frac{\exppq-
\exppp}{\tan\frac{1}{2}(q-p)} \right)=\\ [2mm]
&=&\frac{1}{4\pi}\intp\frac{\exppq-\exppp\left[1+(im-4it\sin p)sin(q-p)
\right]}{\sin^2\frac{1}{2}(q-p)}\hs\nonumber
\ENA
(see eq. (\ref{anine}) for the definition of $\ec$), one rewrites
(\ref{atwelve}) as
\EQA
d(m,t,p_a)&\longrightarrow&\dc(m,t,p)=\nonumber\\
&=&\exppp+\frac{2}{M}\frac{\partial\hs}{\partial p} \ec (m,t,p)\hs.
\ENA


\end{document}